\documentclass[showpacs,amsmath,amssymb,aps,twocolumn,superscriptaddress,prx]{revtex4-2}
\usepackage{mathtools}
\usepackage{algorithm}
\usepackage{algpseudocode}
\usepackage{multirow}
\usepackage[pdftex]{graphicx} \graphicspath{{}}
\usepackage{grffile} 
\usepackage{comment}
\usepackage{makecell}
\usepackage{braket}
\usepackage{placeins}
\usepackage{diagbox}
\usepackage{bbding}
\usepackage{float}
\usepackage{afterpage}
\usepackage{tikz}
\usepackage{ulem}
\usepackage{dcolumn}
\usepackage{bm}
\usepackage[pdftex,colorlinks=true]{hyperref}
\hypersetup{
	colorlinks=true,
	linkcolor=blue,
	urlcolor=cyan,
}

\usepackage{color}
\definecolor{ForestGreen}{RGB}{34, 139, 34}

\newcommand{\zp}[1]{\textsf{\color{blue}#1}}

\newcommand{\affA}{State Key Laboratory of Artificial Microstructure and Mesoscopic
Physics, School of Physics, Peking University, Beijing 100871, China}
\newcommand{\affB}{Max Planck Institute for the Physics of Complex Systems, N\"othnitzer Str.~38, 01187 Dresden, Germany}
\newcommand{\affC}{Institute for Advanced Study, Tsinghua University, Beijing 100084, People’s Republic of China}
\newcommand{\affD}{Hefei National Research Center for Physical Sciences at the Microscale and School of Physical Sciences,
University of Science and Technology of China, Hefei 230026, China}

\begin{document}

\title{Protecting Quantum Simulations of Lattice Gauge Theories \\ 
through Engineered Emergent Hierarchical Symmetries}

\author{Zhanpeng Fu}
 \affiliation{\affC}
\author{Wei Zheng}
 \affiliation{\affD}
\author{Roderich Moessner}
 \affiliation{\affB} 
\author{Marin Bukov}
 \affiliation{\affB} 
\author{Hongzheng Zhao}
 \email{hzhao@pku.edu.cn}
 \affiliation{\affA}

\date{\today}

\begin{abstract}
We present a strategy for the quantum simulation of many-body lattice models with constrained Hilbert spaces. We focus on
lattice gauge theories (LGTs), which underlie a wide range of phenomena in particle physics, condensed matter, and quantum information. In present-day quantum computing platforms, perfect restrictions of the Hilbert space to the desired gauge sectors are beyond reach: for LGTs, violations of the local constraint are unavoidable,
posing a formidable challenge for the emulation of the underlying physics. 
Here, we develop a Floquet-engineering framework that restructures departures from a target sector
such that a series of emergent local symmetries occurs hierarchically in time and in a controllable way. This leads to a set of approximate dynamical selection rules that strongly restrict inter-sector couplings, resulting in a pronounced, symmetry-controlled hierarchy of lifetimes for the state population to spread among sectors. 
Concretely, this protects $U(1)$ LGTs against violations of the {defining} local symmetry. While some sectors remain very long-lived, others are destabilized on shorter timescales. We numerically verify our theory for the one-dimensional $U(1)$ quantum link model.  In addition, we reveal that `defects', whose movement accounts for violations of the gauge constraint, are kinetically constrained,
becoming mobile only through the assistance of intra-sector dynamics, which we describe using an effective quantum marble model.
Our results can thus be used to extend the lifetime, in the spirit of passive error correction, of quantum simulations of complex many-body problems when emergent or desired local symmetries are only implemented approximately.
\end{abstract}

\maketitle
\maketitle
\let\oldaddcontentsline\addcontentsline
\renewcommand{\addcontentsline}[3]{}

\section{Introduction}
Lattice gauge theories (LGTs) play a central role in physics ranging from particle physics~\cite{kogut1979introduction} to models for low-energy effective description in condensed matter systems~\cite{wegner1971duality, anderson1987resonating, kivelson1987topology, zhang1989effective,castelnovo2008magnetic, ross2011quantum}, and quantum error correction~\cite{kitaev2003fault, bombin2006topological, iqbal2025qutrit, bravyi2010topological, lidar2013quantum}. 

A key feature of LGTs~\cite{kogut1979introduction} is that they exhibit local conservation laws, such as  Gauss' laws of electromagnetism, which in turn generate gauge transformations and are encoded by {\it local}  symmetries. 
Hence, the entire many-body Hilbert space splits into multiple decoupled sectors. In elementary particle physics, the sectors violating Gauss' law are unphysical. By contrast, in condensed-matter settings, gauge theories tend to be {\it emergent}, so that the gauge constraint arises not from `fundamental' physics but rather from the structure of the Hamiltonian. This has two basic consequences. 

First, which sectors then constitute the low-energy space, is a matter of detail, in the sense that changes to the Hamiltonian, such as the addition of an applied field, can change the nature of the low-energy sectors. For example, in frustrated systems where gauge theories arise as effective low-energy descriptions, the ground states, e.g.\ of spin liquid nature, may reside in different gauge sectors~\cite{yao2007exact,cassella2023exact}. Moreover, charge configurations corresponding to different sectors can act as an effective internal quenched disorder, which can suppress transport and lead to disorder-free localization~\cite{smith2017disorder}.

Second, even sectors that violate Gauss' law can occur, and are in that sense not unphysical. In a given realization, their existence may be a nuisance to be avoided~\footnote{Indeed, there is some non-uniformity with respect to nomenclature here. The sectors obeying Gauss' law may or may not include gauge charges as sources and sinks of gauge flux. These charges may or may not be mobile. In cases where Gauss' law is violated, one can retain the language of charges by assigning so-called background charges, the presence of which restores the bookkeeping of Gauss' law.}.  
This situation is particularly relevant in the context of 
recent rapid developments in synthetic quantum simulators, which aim to achieve controlled experimental access to specific desired sectors and their unitary dynamics, thereby providing a natural platform to explore the aforementioned effects~\cite{cheng2024emergent,halimeh2025cold,banerjee2013atomic, zohar2013cold, tagliacozzo2013simulation, zohar2015quantum, martinez2016real, schweizer2019floquet, 
surace2020lattice, luo2021gauge,zhou2022thermalization, mueller2022thermalization,buvca2023unified,homeier2023realistic,SunEngineeringCLS2023,jin2023fractionalized,domanti2024floquet, meth2025simulating, hanada2025gauge,will2025probing,mueller2025quantum,hu2025many,datla2026statistical}.  

In the quantum simulation setting, the aim is to restrict dynamics to occur within the set of sectors deemed physical for the study in question. 
In practice, time evolution may include processes connecting sectors that would be entirely disjoint for a perfectly implemented gauge theory. This can happen because the quantum simulation of LGTs normally requires implementing multi-body interactions~\cite{zohar2015quantum,feldmeier2024quantum,geim2026engineeringquantumcriticalitydynamics}. One way to achieve this is by properly introducing energetic penalties that separate different sectors in energy, or through  Floquet engineering using time-periodic (Floquet) driving~\cite{martinez2016real, schweizer2019floquet, domanti2024floquet, meth2025simulating, homeier2023realistic}. 
Such control schemes, together with the associated experimental imperfections, inevitably introduce perturbations that violate the local gauge symmetry.  
On the longest timescales, when the real system with its imperfectly realized gauge symmetry experiences ergodic evolution over all sectors, the LGT description is no longer applicable.

For these reasons, much effort has been invested to enhance the stability of quantum simulation of LGTs.
However, it remains largely unexplored how different sectors respond to perturbations to the pristine gauge theory, and how inter-sector couplings interplay with intra-sector dynamics.
Resolving these questions is crucial for enabling reliable quantum simulations of LGTs and, more fundamentally, for understanding how ergodicity is restored when local conservation laws are weakly broken.

\begin{figure*}[t]
    \centering
    \includegraphics[width=1.0\linewidth]{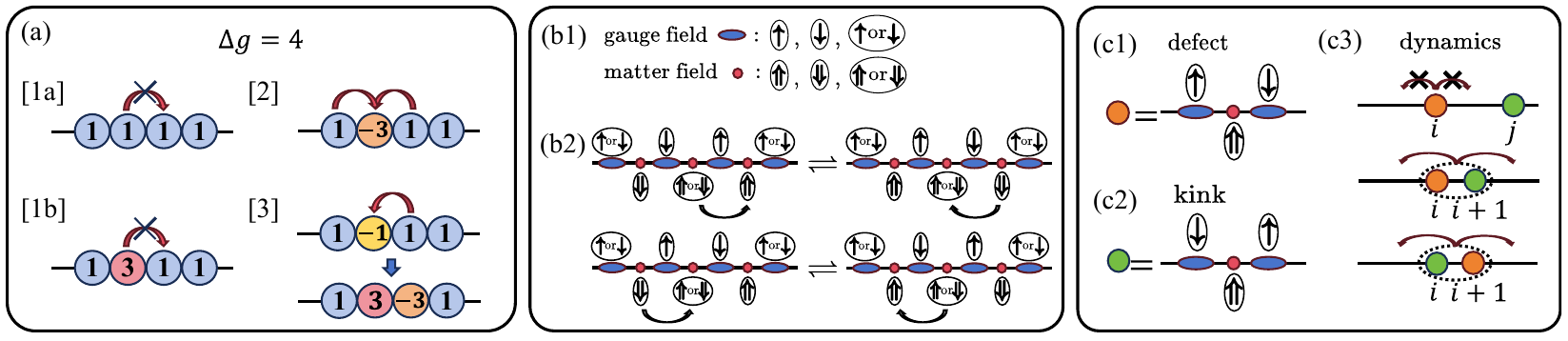}
    \caption{
    Graphic representation of the local symmetry violation dynamics. 
    (a) Defect configurations and their dynamics in the system. The first column corresponds to cases 1(a) and 1(b) and their local $\mathbb{Z}_2$ symmetry-induced constrained dynamics. The second column corresponds to cases 2 and 3, which can be suppressed by Floquet engineering.
    (b1) shows symbols that represent the basis consisting of the eigenstates of $\sigma_j^z$ and $\tau^z_{j,j+1}$ for the matter field and gauge field. (b2) represents the constrained dynamics caused by a first-order perturbation which breaks local $U(1)$ symmetry but preserves local $Z_2\; \times$ global $U(1)$ symmetry. The bottom arrow shows the direction of the local charge transferring. 
    (c1), (c2), (c3) Dynamics of the effective quantum marble model. (c1), (c2) Orange particles represent the "defect" $g_j = -3$ configuration, while green particles are "the kink" configuration with a spin-up matter field in the defect-free sector $g_j=1$. (c3) Defects can only move when kinks come to be the neighbors of defects. Defect-kink dimers can hop freely together, but kinks can break away from the dimer at any time.   
    }
    \label{fig:graprep}
\end{figure*}

In this work, we address these challenges and develop a general Floquet-engineering framework to control the inter-sector coupling, thereby
protecting abelian $U(1)$ LGTs against perturbations that violate the defining local symmetry. The key conceptual ingredient is to dynamically restructure these perturbations, such that a series of emergent local symmetries, e.g., $U(1)^\text{local}$ and $\mathbb{Z}^\text{local}_2$, can occur hierarchically in time and in a controllable way. 
This leads to a set of {approximate} dynamical selection rules that strongly restrict the inter-sector couplings, thereby constraining the spreading of 
‘defects’~(Fig.~\ref{fig:graprep} (a)) , which account
for violations of the gauge constraint, and significantly enhancing the robustness of LGTs. 

Importantly, the engineered selection rules lead to a sharp contrast in leakage rates out of the different sectors:
whereas some sectors 
remain long-lived due to an approximate freezing of defect dynamics, others are destabilized on short timescales. This reveals an intrinsic, symmetry-controlled hierarchy in the robustness of LGTs, which, to the best of our knowledge, has not been identified in previous studies.

Since these selection rules arise solely due to the engineered hierarchical symmetry structure, our protection scheme applies to different $U(1)$ LGTs, irrespective of the specific microscopic details of the underlying model.
In addition, the flexibility of quantum simulators in preparing distinct sectors provides a unique opportunity to directly observe and quantify sector-dependent stability effects in experiments.

For concreteness, we focus on the paradigmatic one-dimensional $U(1)$ quantum link model, with spin-$1/2$ gauge degrees of freedom, although our theory also applies to gauge fields with larger local Hilbert spaces and in higher dimensions. We explicitly elaborate on the selection rules and numerically verify the slow violation of LGTs across different sectors. Crucially, using a fast drive, we show that the lifetime of the local conservation law is algebraically tunable as a function of the drive frequency, which we justify using perturbative theory. 

Further, we identify an exact
mapping from the perturbed quantum link model, and introduce a simplified effective description -- the quantum marble model
(QMM). The QMM reveals that the defect is kinetically constrained and becomes mobile only under intra-sector dynamics.
As a result, the QMM goes beyond symmetry-based selection rules and provides a microscopic description of the interplay between inter-sector coupling and intra-sector coherent dynamics.

Finally, we also find that the presence or absence of approximate many-body degeneracies in the QMM spectrum plays a decisive role in controlling the spreading of defects; we analyze this mechanism using perturbation theory, which yields quantitative predictions for the timescale at which inter-sector coupling becomes appreciable, in excellent agreement with numerical simulations.

The content of this paper is organized as follows. In Sec.~\ref{sec: model}, we first introduce the quantum link model and illustrate possible experimentally relevant perturbations that violate the local conservation law. For pedagogical reasons, in Sec.~\ref{sec.selectiontrule}, we begin with the non-driven quantum link model perturbed by $\mathbb{Z}^\text{local}_2$-preserving terms and elaborate on the selection rules. This precedes the discussion of generic perturbation and the Floquet engineering protocol using single-bond pulses that are already experimentally accessible in Sec.~\ref{sec:HSB}, where we also introduce the QMM.  
We numerically verify our theory in Sec.~\ref{sec: num_res} and explicitly demonstrate the sector-dependent stability. 
In Sec.~\ref{sec:pert_scaling}, we 
show the details of the perturbation theory and illustrate how the approximate degeneracy speeds up the spreading of the defects. We discuss the generalization of our analysis to larger spin-$S$ gauge fields and illustrate open questions in Sec.~\ref{sec: discuss}.

\section{Quantum Simulation of $U(1)$ Lattice Gauge Theory}%
\label{sec: model}
We consider the quantum link model -- a one-dimensional $U(1)$ lattice gauge theory described by
\begin{equation}
\label{eq:H_LMG}
H_{\text{LGT}} = \sum_{j=1}^L \sigma_j^+\tau_{j,j+1}^+\sigma_{j+1}^-+h.c.,
\end{equation}
where $\sigma_j$ is the matter field and $\tau_{j,j+1}$ -- the gauge field. We focus on spin-$1/2$ gauge fields for simplicity: both $\sigma_j$ and $\tau_{j,j+1}$  are represented by Pauli matrices; below, we also discuss higher-spin generalizations of the gauge field $\tau_{j,j+1}$. Multi-body interactions in the Hamiltonian $H_{\text{LGT}}$ can be experimentally implemented on quantum simulators using either analog~\cite{yang2020observation,mark2025observation} or digital methods~\cite{iqbal2024non,mildenberger2025confinement}.

The Hamiltonian $H_{\text{LGT}}$ is invariant under local $U(1)$ gauge transformations (henceforth denoted $U(1)^\text{local}$ generated by~\footnote{Concretely, we have $ [W_j,H_{\text{LGT}}] = 0$, with  $W_j = e^{-i\theta_jG_j}$ for any $j$ and an arbitrary angle $\theta_j$.} 
\begin{equation}
\label{eq:Gauss_law}
    G_j = \tau_{j,j+1}^z - \tau_{j-1, j}^z - \sigma_j^z.
\end{equation}
This leads to a set of mutually commuting and locally conserved charges, with eigenvalues $g_j\in\{\pm 1, \pm 3\}$.
A configuration $\textbf{g} =
(g_1, g_2, \dots , g_L)$ over the entire lattice defines a sector, and different sectors are decoupled.

We represent the state of the gauge field $\tau_{j,j+1}$ on the $(j,j+1)$-links by upper arrows $^\uparrow, ^\downarrow$ and the state of the matter field $\sigma_j$ on the sites $j$ by lower double arrows $_\Uparrow, _\Downarrow$. Then, using Eq.~\eqref{eq:Gauss_law},  $g_j{=}1$ corresponds to states that look locally like $|\cdots \;^{\downarrow_{j-1,j}} \; _{\Downarrow_j} \;^{\downarrow_{j,j+1}} \cdots \rangle$
, $|\cdots \;^{\uparrow_{j-1,j}} \; _{\Downarrow_j} \;^{\uparrow_{j,j+1}} \cdots \rangle$, or $|\cdots \;^{\downarrow_{j-1,j}} \; _{\Uparrow_j} \;^{\uparrow_{j,j+1}} \cdots \rangle$.

In quantum simulation experiments, local symmetry $U(1)^\text{local}$ is usually violated by the presence of additional unwanted perturbations~\cite{schweizer2019floquet, zhou2022thermalization}. We model the experimental system using the Hamiltonian 
\begin{eqnarray}
\label{eq:lab}
    H &=& J H_{\text{LGT}} + KH_1  + h H_0, \\
    H_1 &=& \sum_j \sigma_j^+\tau_{j,j+1}^x\sigma_{j+1}^-+h.c., 
     \nonumber\\
     H_0 &=& \sum_j  \tau_{j,j+1}^x+\varepsilon_1\sigma_j^z\tau_{j,j+1}^x+\varepsilon_2\tau_{j,j+1}^x\sigma_{j+1}^z , \nonumber
\end{eqnarray}
where $H_{0,1}$ describe
terms that violate the target local symmetry, and $h,J,K$ set the corresponding energy scales; $\varepsilon_{1,2}$ are dimensionless relative strength parameters for the various interactions between matter and gauge fields. Note that $H_1$ preserves both the $\mathbb{Z}^\text{local}_2$ symmetry defined by 
\begin{equation}
    S_j {=} i e^{-i\frac{\pi}{2} G_j} {=} - \tau_{j-1,j}^z\, \sigma_{j}^z\, \tau_{j,j+1}^z \, ,
    \label{eq:Sj}
\end{equation} 
and hence $H_1{=}S_j^{\dagger} H_1S_j$, as well as the global $U(1)$ symmetry: $\sum_j [G_j, H_1] = 0$.
Other generic perturbations are captured by $H_0$ devoid of any {local} symmetry structure. 
For simplicity, we consider only terms in $H_0$ that preserve the global $U(1)$ symmetry corresponding to particle number conservation, which is commonly satisfied in quantum simulation platforms~\footnote{In the presence of additional terms that break global $U(1)$ symmetry, it is also feasible to achieve HSB structure with potentially more complex driving protocols.}.

Synthetic quantum simulators have achieved controlled experimental access to individual sectors by preparing the initial state with different $\textbf{g}$ configurations. While elementary particle physics focuses on the sector $\textbf{g}=\{1,-1,1,-1\dots\}$ where Gauss' law is obeyed, in this study we will not be limited to this specific choice. 
Instead, we will show that different sectors exhibit sharply different levels of robustness against perturbations that violate the local symmetry.
For this purpose, we find that it is convenient to start with the sector with $g_j{=}1,\forall j$, which we refer to as the \textit{defect-free} sector, $\text{GS}_0$. Other sectors allow $g_j \ne 1$ on site $j$, and we refer to this as a {\it defect}. 

Our aim is to restrict quantum dynamics to occur within a given sector. However, when the perturbation shown in Eq.~\eqref{eq:lab} is present, two dynamical consequences will occur: (i) quench dynamics under the generic perturbations described by $H$ cause the rapid spreading of defects and couples different sectors, labeled by the spatial rearrangement of $\textbf{g}$ along the lattice. 
Furthermore, (ii), $U(1)^\text{local}$ symmetry-violating perturbations due to $H_0$ or $H_1$ in the quench dynamics will themselves produce defects.
These violations impose severe limitations on the time window available for investigating the physics of the $U(1)$ LGT model $H_\text{LGT}$ in experiment.

{
The key conceptual contribution of this work is the design and construction of a Floquet protocol that protects the target $H_{\text{LGT}}$ dynamics on a parametrically tunable time scale, by exploiting the symmetry structure of the problem. 
Hence, we fully address issues (i) and (ii). This is achieved by imprinting into the time evolution a tailored hierarchical symmetry structure \footnote{The symmetry group here satisfies $U^\text{local}(1) \supset \mathbb{Z}^\text{local}_2 \times U(1)^\text{global}\supset E$. In Floquet dynamics, due to the structure of the effective Hamiltonian, the system sequentially enters prethermal plateaus that preserve the corresponding symmetry along the direction of the arrow.}
\begin{equation}
\label{eq:hsb_structure}
    U(1)^\text{local} \to \mathbb{Z}^\text{local}_2 \times U(1)^\text{global}\to E ,
\end{equation}
with $E$ denoting the trivial group. The emergent symmetry $\mathbb{Z}^\text{local}_2 \times U(1)^\text{global}$ introduces a set of selection rules which significantly slow down the spreading of existing defects (i), and forbid certain types of production channels of new defects (ii).
Hence, the resulting protocol protects the target $H_{\text{LGT}}$-dynamics for a parametrically long timescale.
}

\section{Selection Rules}%
\label{sec.selectiontrule}

For pedagogical purposes, before presenting the physics of our tailored dynamical protocol, let us first focus on the Hamiltonian Eq.~\eqref{eq:lab} with $h=0$, and the underlying physics. 

{We now show} that the Hamiltonian $H(h{=}0)$ exhibits a $\mathbb{Z}^\text{local}_2 \times U(1)^\text{global}$ sub-symmetry, which leads to a set of selection rules that protect the target $U(1)^\text{local}$ symmetry. 
These selection rules only rely on the symmetry structure of the perturbation, and hence they do not depend on the specific form of the perturbation. They also apply without an external drive, although, as we show later, Floquet engineering can further enhance the performance of the protection scheme. 

To explain the essence of the physics behind the selection rules, let us focus solely on the dynamics of the local charge $g_i$; we present the complete derivation involving the elementary matter and gauge fields in the supplemental material (SM).

Given a fixed site $j$, notice first that $\mathbb{Z}^\text{local}_2$ preserving operations can induce transitions only within one of the two sector pairs 
\begin{subequations}
    \begin{equation}
    \label{eq.onsitesrule}
        g_j = (-3,1) \qquad \text{and} \qquad g_j=(-1, 3)\, ,
    \end{equation}
because spin configurations within each pair possess the same eigenvalues of the $S_j$ operators defining the $\mathbb{Z}^\text{local}_2$ symmetry, cf.~Eq.~\eqref{eq:Sj}.

Second, for two neighboring sites $j$ and $j+1$, the $U(1)^\text{global}$ symmetry ensures the conservation of the total charge  $\sum_jg_j$, such that only three possibilities for the dynamics are allowed: 
    \begin{eqnarray}
    \label{eq.twositesrule}
        \{g_j,g_{j+1}\}&=&\{-1,3\}{\leftrightarrow}\{3,-1\}\quad  \text{or}\quad  \{-3,1\}{\leftrightarrow}\{1,-3\}\nonumber\\ 
        &&\text{or} \quad \{-1,1\}{\leftrightarrow}\{3, -3\},
    \end{eqnarray}
\end{subequations}
so that the charge exchanged is $\Delta g=4$, and the sum $g_j{+}g_{j+1}$ is preserved by this local process. In particular, no other local processes are allowed. Therefore, the spreading of defects is kinetically constrained.

The arising dynamical consequences can be better understood by focusing on configurations with a low density of defects,
where for most of the site indices $j$ we have $g_j=1$. 
Using the selection rules, the behavior of the most probable defect configurations in the system can be classified into four processes due to the selection rule in Eq.~\eqref{eq.twositesrule}, cf.~Fig~\ref{fig:graprep}(a): 
\begin{enumerate}
    \item[[1a\!\!]]: no defects can be created from a defect-free configuration with $\{g_i\}{=}\{\dots,1,1,1\dots\}$;
    \item[[1b\!\!]]: defects with $g_j{=}3$, e.g., $\{g_i\}{=}\{\dots,1,3,1\dots\}$, will be frozen;
    \item[[2\!\!]]: a defect with $g_j{=}{-}3$ can move freely only if its neighbors are defect-free, e.g., $\{g_i\}{=}\{\dots, 1,-3,1,\dots\}\to\{\dots,1, 1,-3,\dots\}$;
    \item[[3\!\!]]: a defect with $g_i{=}{-}1$ can generate a pair of defects, $\{g_i\}{=}\{\dots,1,-1,1,\dots\} \to \{g_i\}{=}\{\dots,1,3,-3,1,\dots\}$, but only one of them can freely propagate.
\end{enumerate}
Therefore, these selection rules create additional stability in the defect-free sector. 
The charge dynamics are dominated by the spreading of the $g_j{=}{-}3$ defects according to cases 2 and 3.
In the following, we propose a Floquet protocol, which realizes the hierarchical symmetry structure $U(1)^\text{local} \to \mathbb{Z}^\text{local}_2 \times U(1)^\text{global}$ to further suppress the spreading and enhance the protection. 

Even though for generic perturbations with $h\neq0$ the above selection rules are no longer exact, the Floquet-engineered hierarchical symmetry structure still ensures their validity for parametrically long times, as we demonstrate explicitly in Sec.~\ref{sec: num_res}.

\section{Dynamical Protection via Engineered Hierarchical Symmetries}%
\label{sec:HSB}

We first discuss the Hamiltonian $H(t)$ at $h=0$ in Sec.~\ref{subsec:h=0}. We derive an effective model to describe the defect dynamics in Sec.~\ref{sec.QMM}, and generalize the results to $h\neq 0$ in Sec.~\ref{subsec:genericperturbation}.

\subsection{Floquet protocol for $h{=}0$}
\label{subsec:h=0}

Let us define the single-bond pulse operators $P_{\tau}^z = \exp(-i\frac{\pi}{2}\sum_j\tau_{j,j+1}^z)$, $P_{\tau}^x = \exp(-i\frac{\pi}{2}\sum_j\tau_{j,j+1}^x)$, realized experimentally by applying an external field on gauge links $\tau$~\cite{domanti2024floquet, mi2022time, beatrez2023critical, zhang2022digital, dumitrescu2022dynamical}. 
A tailored design of a sequence of pulses can effectively echo out the unwanted symmetry-breaking perturbation~\cite{fu2024engineering}.

Consider the following evolution operator over one period $T_F = (2+J/K)T$ 
\begin{equation}
\label{eq:dirve_simple}
    U_F {=} \exp\left(-iH\frac{J{+}K}{K}T\right)P_{\tau}^zP_{\tau}^x\exp\left(-iHT\right)P_{\tau}^{x\dagger}P_{\tau}^{z\dagger}\ ,
\end{equation}
where $T$ is the {duration} between the single-bond pulses, and the Hamiltonian $H=JH_\text{LGT}+KH_1+hH_0$, cf.~Eq.~\eqref{eq:lab}. 
To understand the resulting dynamics, one can define a static effective Hamiltonian $Q_F$ through the relation $U_F{\equiv} e^{-iQ_FT_F}$. In the high-frequency regime where the drive frequency $\omega_F = 2\pi/T_F$ is larger than the local energy scales of $H$, one can perturbatively determine
$Q_F{=}\sum_m Q_F^{(m)}$, with $Q_F^{(m)}{\sim}T_F^m$, and its truncation at a finite order can be used to approximate the stroboscopic time evolution of the system.

The sequential application of $P_{\tau}^{x/z}$ effectively echoes out the 
{$\mathbb{Z}^\text{local}_2 \times U(1)^\text{global}$} symmetry-violating perturbation $H_1$ at the order $\mathcal{O}(T_F)$; consequently, the dynamics to the order $\mathcal{O}(T_F^2)$ can be captured by $Q_F^{(0)}+Q_F^{(1)}$ with 
\begin{eqnarray}
Q_F^{(0)}&{=}&JH_{\text{LGT}},\nonumber\\
Q_F^{(1)}&{=}&i\lambda_0[H_{\text{LGT}}, H_1],\qquad \lambda_0 {=} \frac{KJ(J{+}K)}{2(2K{+}J)}T_F,
\label{eq:eff_1}
\end{eqnarray}
{realizing the $U(1)^\text{local} \to \mathbb{Z}^\text{local}_2 \times U(1)^\text{global}$ hierarchical symmetry structure~\eqref{eq:hsb_structure}.}
The absence of $H_1$ in the zeroth-order effective Hamiltonian $Q_F^{(0)}$ in Eq.~\eqref{eq:eff_1} causes the first-order terms $Q_F^{(1)}$ to dominate the defect dynamics.
For the sake of clarity and simplicity, we show the explicit form {of the effective Hamiltonian Eq.~\eqref{eq:eff_1}} in~\ref{sec:pert}.

\begin{figure*}[t]
    \centering
    \includegraphics[width=1.0\linewidth]{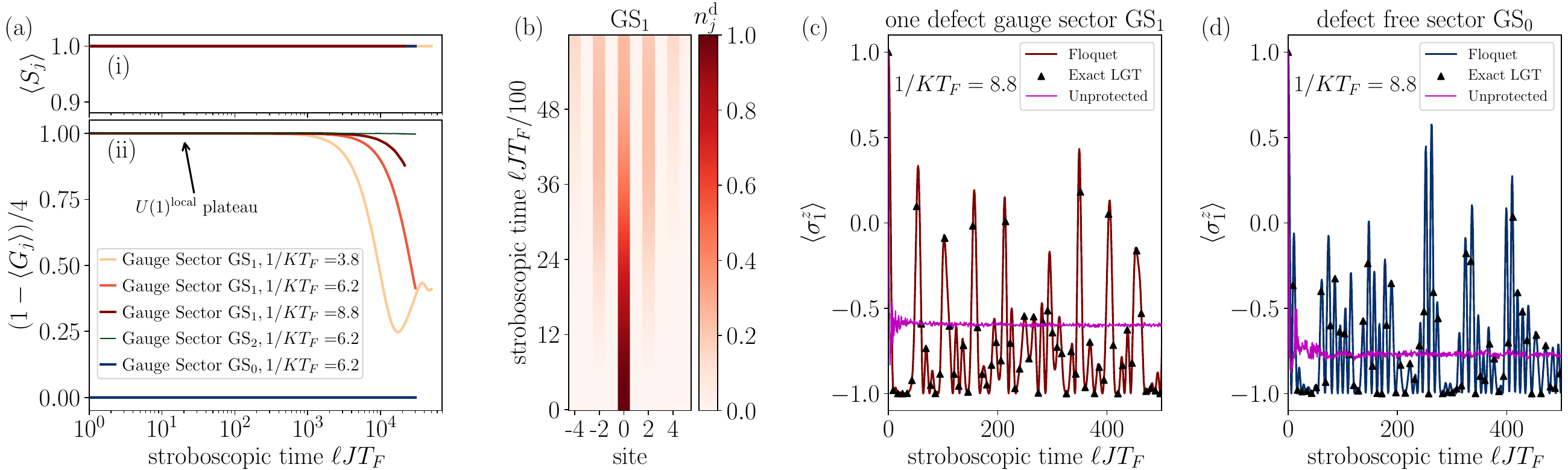}
    \caption{Dynamical protection of matter and gauge field dynamics. (a) Dynamics of the charge for the initial sector with defect $\text{GS}_1:\{g_j\} = \{-3, 1, 1, \cdots, 1\}$ (red, PBC) and $\text{GS}_2:\{g_j\} = \{-3, 1, 1,1,1,-3,1,1,1,1\}$ (green, PBC) at site $j=0$, and the defect-free sector 
    $\text{GS}_0:\{g_j\} = \{1, 1, 1, \cdots, 1\}$ (blue, OBC) at site $j=1$. The inserted figure: dynamics of (i) $\mathbb{Z}_2^\text{local}$ symmetry and (ii) the $U(1)^\text{local}$ symmetry violation for the local charge. The prethermal lifetime of (i) is much longer than the lifetime of (ii).
    For sector $\text{GS}_0$, we show the dynamics of the charge at site $j=1$, which is frozen until high-order perturbations take effect after a long period of time. For the sector with a defect GS$_1$, we show the dynamics at site $j=0$, where the defect spreads slowly and the lifetime of the prethermal plateau can be controlled by the driving frequency.
    (b) shows the spatio-temporal distribution of defect density $n^{\text{d}}_j = (1-\langle G_j\rangle)/4$ for the sector $\{g_j\} = \{-3, 1, 1, \cdots, 1\}$ with the driving frequency $1/KT_F = 3.8$. Despite the lower driving frequency, defects remain largely localized to their initial positions for a long time. Moreover, due to the structure of the effective Hamiltonian, the motion of defects exhibits a striped pattern.
    (c,d) Comparison between the dynamics of the matter field in the system under the new driving protocol shown in \ref{sec:newprotocol} (red and blue lines), the exact LGT (triangle), and the unprotected quench dynamic of $H$ in Eq.~\eqref{eq:lab} (purple line) in the two sectors $\text{GS}_1$ (b) and $\text{GS}_0$ (c). We can observe that the system's dynamics can be well described by LGT under our driving protocol. However, under the same perturbation strength, the system's quench dynamics without driving will rapidly deviate from the target LGT dynamics.
    The parameters in the simulation are $J = 1.0$, $K = 4.0,  h = 0.5, \epsilon_1 = \epsilon_2 = 1.0$. System size is $L = 10$.
    }
    \label{fig:floq}
\end{figure*}

Importantly, we note that $Q_F^{(1)}$ generates a new type of kinetic constraint in addition to the ones induced by the selection rules shown in Sec.~\ref{sec.selectiontrule}.
As shown in Fig.~\ref{fig:graprep} (b2), the {gauge} charge transfer between site $j$ and site $j\pm1$ can only occur when there are opposite matter and gauge spins on sites $j\pm 1$ and the links between $j$ and $j\pm 1$. 
This is equivalent to inhibiting direct charge exchange between adjacent lattice sites (process [2] in Sec.~\ref{sec.selectiontrule}). 
Hence, the resulting constrained dynamics correspond to defects $g_j=-3:\; |\cdots \; ^\uparrow \; _\Uparrow \;^\downarrow \cdots\rangle $ that can only move if they are adjacent to a specific configuration in $g_i=1: \; |\cdots\;^\downarrow \; _\Uparrow \;^\uparrow \cdots \rangle$, where single-arrows denote the state of the gauge field. We refer to the latter configuration as \textit{kink} because the spin-up matter field $\Uparrow$ causes the spins of the gauge fields on both sides to be opposite, cf.~Fig.~\ref{fig:graprep}(c2).

\subsection{Quantum marble model (QMM)}%
\label{sec.QMM}

We now identify an exact mapping from the Hamiltonian $Q_F^{(0)}{+}Q_F^{(1)}$ 
to a simplified effective model for configurations with one defect and one kink. Because the motion of a defect is allowed only when it collides with a kink, we refer to this effective model as the Quantum Marble model (QMM)~\footnote{For simplicity, here we focus on systems with open boundary conditions (OBC); generalization for periodic boundary conditions can be found in the Supplementary Material.}.
{As we demonstrate below, the QMM not only provides a clearer illustration of the 
kinetically
constrained dynamics of defects, but also enables numerical simulations of defect-spreading dynamics in system sizes that would otherwise be unreachable using exact simulation of the entire Hilbert space.} 
{Crucially, an explicit analysis of the QMM spectral degeneracy, as well as its lifting by higher-order (in $T_F$) perturbations, directly determines the stability of the target LGT dynamics.}

Let us first elaborate on the construction of the QMM. We note
that the relative positions of the defect and the kink in an open chain remain unchanged as time evolves: the defect can either be on the left (case 1, cf.~Eq.~\eqref{eq:mapping}) or on the right (case 2) hand side of the kink. Let us therefore define the creation operators $\kappa_j^{\dagger}$ and $\Delta_j^{\dagger}$ as
\begin{eqnarray}
\label{eq:mapping}
\text{case 1:}\quad
\kappa_j^{\dagger} &=& \prod_{k=j}^{L-1}\tau_{k,k+1}^+\sigma_j^+,\quad \Delta_j^{\dagger} = \prod_{k=0}^{j}\tau_{k-1, k}^+\sigma_j^+,\\
\text{case 2:}\quad \kappa_j^{\dagger} &=& \prod_{k=0}^{j}\tau_{k,k+1}^-\sigma_j^+,\quad \Delta_j^{\dagger} = \prod_{k=j}^{L-1}\tau_{k,k+1}^-\sigma_j^+,\nonumber
\end{eqnarray}
where $\Delta_j^{\dagger}$ creates a defect with $g_j{=}{-}3$ on top of a defect-free configuration, while $\kappa_j^{\dagger}$ creates a kink in the defect-free subspace. In both cases, the operators for kink and defect commute, i.e., $[\kappa_i^{(\dagger)}, \Delta_j^{(\dagger)}] = 0$.
Specifically, we choose the state $|\;^{\downarrow}\;_{\Downarrow}\;^{\downarrow}\cdots\rangle$ as the vacuum state for case 1 and the state $|\;^{\uparrow}\;_{\Downarrow}\;^{\uparrow}\cdots\rangle$ for case 2, cf.~Fig~\ref{fig:graprep}(c1) and (c2).
Recall that the operator $\sigma_j^{+}$ flips up a spin at the site of the matter field, while the operator $\tau_{k,k+1}^{+/-}$ flips up or flips down a spin at the site of the gauge field to keep the state in the fixed sector. Hence,
$\kappa_j^{\dagger}$ and $\Delta_j^{\dagger}$ obey a mutual hard-core condition, i.e.,  $(\kappa_j^{\dagger})^2=(\Delta_j^{\dagger})^2=\kappa_j^{\dagger}\Delta_j^{\dagger}=0$.

One can also identify the mapping between the local observables in the original LGT model and in the QMM. 
The density of kinks $n_j^{\text{k}}$ (equivalently the matter field density in the exact LGT), and the defect density for $g_j{=}-3$, $n^{\text{d}}_j$ (local charge violation) 
can be mapped as
\begin{eqnarray}
    &&n_j^{\text{k}}=\left\langle (1+\sigma_j^z)/2\right\rangle|_{g_j=1}=\langle \kappa^{\dagger}_j\kappa_j\rangle, \nonumber\\
    &&n^{\text{d}}_j = (1{-}\left\langle G_j\right\rangle)/4 = \langle \Delta_j^{\dagger}\Delta_j\rangle,
\end{eqnarray}
where $\langle \cdot \rangle|_{g_j=1}$
suggests that the expectation value is computed using configurations with the charge $g_j=1$.

As detailed in~\ref{sec:QMM}, $H_{\text{LGT}}$ and $H_1$ can be exactly represented using these operators as
\begin{eqnarray}
    H_{\text{LGT}} &=& \sum_j\kappa_{j}^{\dagger}\kappa_{j+1}+h.c., \nonumber\\
    H_1 &=& \sum_j\kappa_{j}^{\dagger}\kappa_{j+1}+\Delta_j^{\dagger}\Delta_{j+1}+h.c.\, ,
\end{eqnarray}
suggesting that defects and kinks can hop freely on the lattice. Notice that it is the defect hopping, $\Delta_j^{\dagger}\Delta_{j+1}$, that leads to the spreading out of the violation of the local charge conservation.
The QMM Hamiltonian is then obtained from the effective Hamiltonian $Q_F^{(0)}+Q_F^{(1)}$, and reads as
\begin{eqnarray}
\label{eq.QMMeffective}
    H_{\text{QMM}} 
    = \sum_j&&\left(J\kappa_{j+1}^\dagger \kappa_j+i\lambda_0\Delta_j^{\dagger}\kappa_{j+1}^{\dagger}\Delta_{j+1}\kappa_{j+2}\right.\\ &&-i\left.\lambda_0\kappa_j^{\dagger}\Delta_{j+1}^{\dagger}\kappa_{j+1}\Delta_{j+2}+h.c.\right).\nonumber
\end{eqnarray}

We visualize the QMM dynamics in Fig~\ref{fig:graprep}(c3): a defect (orange circle) can only move after coupling with a kink (green circle), whereas a kink can be detached from a defect at any time and move freely in the system via $Q_F^{(0)}$. Since the dynamics of the defect are highly dependent on its {collision} with the kink, it is intuitively clear that defect spreading across the lattice will be highly suppressed.

Although the mapping~\eqref{eq:mapping} is exact only for a single defect and a single kink, we find that the QMM can still exactly capture the dominant dynamics for more complex configurations. In~\ref{sec:QMM}, we show the general mapping that can map the state in the sector $\{g_j=-3,1\}$ to the configuration with multiple defects and kinks. 

We highlight that the mapping to the QMM effectively reduces the original many-body system to a few-body system, resulting in an exponential reduction in the effective Hilbert space dimension. 
{Indeed, for a well-prepared initial state {(i.e., a state with low defect density)}, the dimension of its Hilbert space is $\mathcal{O}(L^{N})$, where $N$ is the total number of defects and kinks, $L$ is the system size, and $N\ll L$.}
This should be contrasted with the entire Hilbert space of an exponentially large dimension $4^{L}$. 
This reduction naturally allows us to investigate the stability of LGT systems in larger systems.

\subsection{General case with $h{\neq}0$}%
\label{subsec:genericperturbation}

In practice, $h{\ne}0$ terms will be present in experiments, and hence the perfect $\mathbb{Z}_2^\text{local}$ symmetry will be lost. Therefore, the protocol in Eq.~\eqref{eq:dirve_simple} needs to be modified such that the hierarchical symmetry-breaking structure~\eqref{eq:hsb_structure} and the dynamical protection it offers, can still be achieved. 

As detailed in~\ref{sec:newprotocol}, it is possible to design a modified Floquet protocol such that the effective Hamiltonian $Q_{F}^{(2)}$ of order $\mathcal{O}(T^2)$ degrades the symmetry group as $U(1)^\text{local} \to \mathbb{Z}^\text{local}_2 \times U(1)^\text{global}$. Essentially, this new protocol is similar to Eq.~\eqref{eq:dirve_simple}, but a few more single-bond pulses are required to realize the echo-out sequence. 

Note that the hierarchical symmetry structure in this case is only approximate since $Q_F^{(2)}$ also involves terms that weakly violate $\mathbb{Z}^{\text{local}}_2$. However, these terms are parametrically small, i.e., of order $\mathcal{O}(h^2/JK)$ compared to other terms of the same order in $T_F$, as shown in ~\ref{sec:newprotocol}, and hence can be safely neglected. 
Under these conditions, Eq.~\eqref{eq:mapping} again provides an effective QMM description, see~\ref{sec:newprotocol} for the explicit form of the QMM Hamiltonian.
Therefore, even for generic symmetry-violating perturbations, defect spreading dynamics can still be kinetically
constrained up to order $\mathcal{O}(T_F^3)$, and hence the target LGT remains robust.

\section{Numerical results}%
\label{sec: num_res}
We numerically verify the dynamical protection of LGT dynamics against generic perturbations, using the Floquet drive in Sec.~\ref{subsec:genericperturbation} for initial states in different sectors. Below, we also explicitly demonstrate the validity of the QMM model. 

\subsection{Protecting lattice gauge theory (LGT) dynamics}
We first start from the simplest defect-free initial configuration with gauge sector $\text{GS}_0: \{g_j\} = \{1, 1, \cdots, 1\}$. As discussed in Sec.~\ref{sec.selectiontrule}, $\mathbb{Z}^\text{local}_2 \times U(1)^\text{global}$ preserving perturbations cannot perturb this 
sector, and hence the dynamics should follow the exact LGT dynamics for very long time. To show this, we consider the initial state $\left|_{\Downarrow}\;^{\downarrow}\;_{\Uparrow}\;^{\uparrow}\;_{\Downarrow}\;^{\uparrow}\;_{\Downarrow}\;^{\uparrow}\cdots_{\Downarrow}\;^{\uparrow}\right\rangle$.
To detect the violation of local symmetries, we measure the normalized $U(1)^\text{local}$ charge $(1-\langle G_j\rangle)/4$ [cf.~Eq.~\eqref{eq:Gauss_law}] and the $\mathbb{Z}_2^\text{local}$ operator $\langle S_j\rangle$  [cf.~Eq.~\eqref{eq:Sj}].
The results are plotted in  Fig.~\ref{fig:floq}(a) (blue, $\text{GS}_0$), and throughout the entire simulation time, these two observables remain close to their initial values, thus validating our dynamical LGT protection scheme.

Now we consider the gauge sector with one defect, $\mathrm{GS}_1$: $\{g_j\} = \{-3,1,1,\cdots\}$ with the initial state $\left|\;^{\uparrow}\;_{\Uparrow}\;^{\downarrow}\;_{\Uparrow}\;^{\uparrow}\;_{\Downarrow}\;^{\uparrow}\cdots_{\Downarrow}\;^{\uparrow}\right\rangle$ that initializes the system with one kink at site 1. In Fig.~\ref{fig:floq}(a), we show the Floquet dynamics of the system (red line). In this case, the defect spreads in space [see spatiotemporal profile of the charge distribution in Fig.~\ref{fig:floq}(b)], and hence the charge $g_j$ is no longer conserved locally. However, as shown in Fig.~\ref{fig:floq}(a) ($\text{GS}_1$), its dynamics can be better constrained by increasing the driving frequency, and the lifetime $\tau$ of the $U(1)^\text{local}$ charge conservation is prolonged accordingly. As shown in Fig.~\ref{fig:scaling}, the corresponding lifetime $\tau$ follows the parametric dependence $T_F^{-2}$, which we explain in Sec.~\ref{sec:pert_scaling}.

The protection of the local symmetry also ensures the accurate evolution of the matter and gauge fields within the corresponding sector.
To show this, we present the matter field dynamics for the sector $\mathrm{GS}_1$ and $\mathrm{GS}_0$ in Fig.~\ref{fig:floq}(c,d), respectively. 
In particular, in our simulation, we consider $K{=}4.0J$, a relatively large perturbation strength of the term violating $U(1)^\text{local}$ but preserving its $\mathbb{Z}_2^\text{local}$ subgroup. Consequently, the unprotected quench dynamics generated by the perturbed Hamiltonian $H$, Eq.~\eqref{eq:lab}, induce large errors (purple line) already at a short $\mathcal{O}(1)$ time scale. In contrast, the Floquet dynamics (red and blue lines) remain stable and precisely match the exact LGT dynamics (black triangle) throughout the entire time window, confirming the validity of our protection scheme. {In Sec.~\ref{sec:physical sector}, we further present simulations in the sector $\{g_j\}=\{1,-1,1,-1,\cdots\}$ where Gauss' law is obeyed and validate our protection scheme.}

\subsection{Verification of quantum marble model (QMM)}
\label{sec.QMMnumerics}

\begin{figure}[t]
    \centering
    \includegraphics[width=1.0\linewidth]{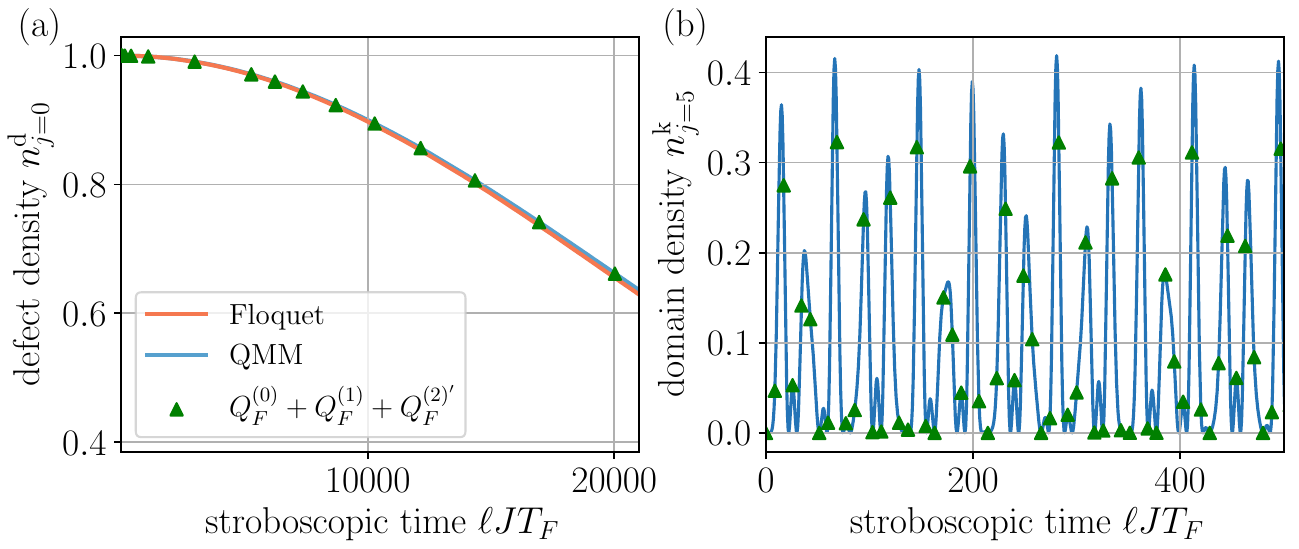}
    \caption{Validation of the quantum marble model(QMM).  
    (a) Evolution of defect density $n^{\text{d}}_j{=}(1{-}\left\langle G_j\right\rangle)/4 \equiv \langle \Delta^{\dagger}_j\Delta_j\rangle$ at site $j{=}0$. Note that, in the second-order effective Hamiltonian, we only focus on terms that preserve $\mathbb{Z}^{\text{local}}_2$, which we denote as $Q_F^{(2)\prime}$. Its explicit expression can be found in ~\ref{sec:newprotocol}.
    (b) Evolution of kink density $n^{\text{k}}_j{=}(1{+}\left\langle \sigma^z_j\right\rangle)/2 \equiv \langle \kappa^{\dagger}_j\kappa_j\rangle$ at site $j{=}5$. 
    Both the violation of the local charge and the kink dynamics can be faithfully captured by QMM throughout an extended time window. The parameters are the same as Fig.~\ref{fig:floq}, with driving frequency $1/KT_F = 6.2$.
    } 
    \label{fig:QMM_dyna}
\end{figure}

We further verify that the QMM described in Sec.~\ref{subsec:genericperturbation} accurately describes the dynamics generated by the effective Hamiltonian and the Floquet protocol. 
In Fig.~\ref{fig:QMM_dyna}, we consider the same state in the sector with one defect, $\mathrm{GS}_1$: $\{g_j\} = \{-3,1,1,\cdots\}$ and the initial state $\left|\;^{\uparrow}\;_{\Uparrow}\;^{\downarrow}\;_{\Uparrow}\;^{\uparrow}\;_{\Downarrow}\;^{\uparrow}\cdots_{\Downarrow}\;^{\uparrow}\right\rangle$. In Fig.~\ref{fig:QMM_dyna}(a), we show the dynamics of the defect located at $j{=}0$, while in Fig.~\ref{fig:QMM_dyna}(b) we show the dynamics of the kink at ($j{=}5$, away from the defect), corresponding to the dynamics of the matter field magnetization in the protected LGT.
The QMM dynamics (green triangles) accurately capture the Floquet (red) and effective Hamiltonian (blue curves) dynamics, validating the applicability of the QMM.

\begin{figure}[t!]
    \centering
    \includegraphics[width=1.0\linewidth]{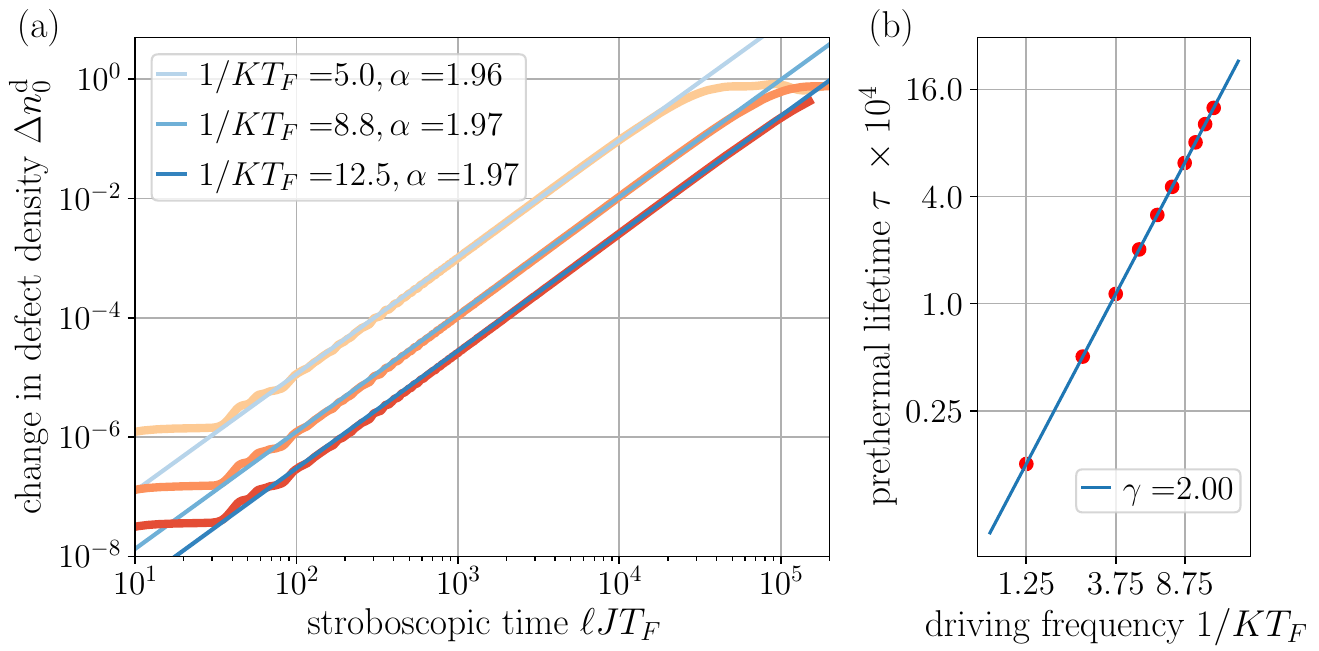}
    \caption{Dynamics of defect and prethermal lifetime scaling versus the driving frequency in QMM within the sector $\text{GS}_1$. The initial state has a defect on site 0 and a kink on site 1, which is equivalent to the initial state of $\text{GS}_1$ in Fig.~\ref{fig:floq}. The dynamics of the QMM is equivalent to the effective Hamiltonian $Q^{(0)}_F+Q_F^{(1)}+Q_F^{(2)\prime}$ corresponding to the original Floquet dynamics. (a) Dynamics of change in defect density on site 0. Over a long time scale, the defect density exhibits an algebraic scaling $\Delta n_j^{\text{d}} \sim t^{\alpha}$ at site $j=0$, where $\alpha$ is close to 2. (b) Scaling of the lifetime of the $U(1)^\text{local}$ prethermal plateau 
    with respect to the driving frequency $\tau\sim T_F^{-\gamma}$. The lifetime is defined as the time for $n_j^{\text{d}}$ decays to $e^{-0.4}$. Combining the results in (a) and (b), the dynamics show that defects will spread into the system with polynomial scaling $\sim (T_F^2t)^2$. The parameter is the same as Fig.~\ref{fig:floq} and the system size is $L = 16$.
    }
    \label{fig:scaling}
\end{figure}

We further use the QMM to numerically simulate a system with 16 matter sites and 16 gauge links, a system size much larger than what can be reached with exact dynamics within the entire Hilbert space. We perform a long-time simulation to track the violation of the {local conservation law} and its dependence on the driving frequency. 
As shown in Fig.~\ref{fig:scaling}(a), we plot the change of defect density $\Delta n_j^{\text{d}}$ at site $j=0$. Note that, under perfect LGT dynamics, $\Delta n_j^{\text{d}}$ always remains zero, while it grows here due to symmetry-violating perturbations and following precisely a $\Delta n_j^{\text{d}}\sim (T_F^2t)^2$ law (blue line in Fig.~\ref{fig:scaling}(a)). Accordingly, it leads to a deviation from the prethermal plateau as seen in Fig.~\ref{fig:floq}; we find that the prethermal lifetime obeys the scaling law $\tau \sim T_F^{-2}$ as shown in Fig.~\ref{fig:scaling}(b).

We now justify these numerical observations using a perturbative analysis in Sec.~\ref{sec:pert_scaling}.

\section{Perturbative analysis of defect dynamics}%
\label{sec:pert_scaling}
The spreading of the defect, and hence violations of the local conservation law of the target LGT, strongly relies on the presence or absence of an approximately degenerate spectrum in the QMM model.
We now demonstrate the origin of the degeneracy, how defects lift the degeneracy, and, in turn, how this degeneracy influences the spreading of defects. Importantly, we show that increasing the number of defects makes the emergence of degenerate structures progressively more difficult, thereby further suppressing defect dynamics.

\begin{figure}[t!]
    \centering
    \includegraphics[width=0.8\linewidth]{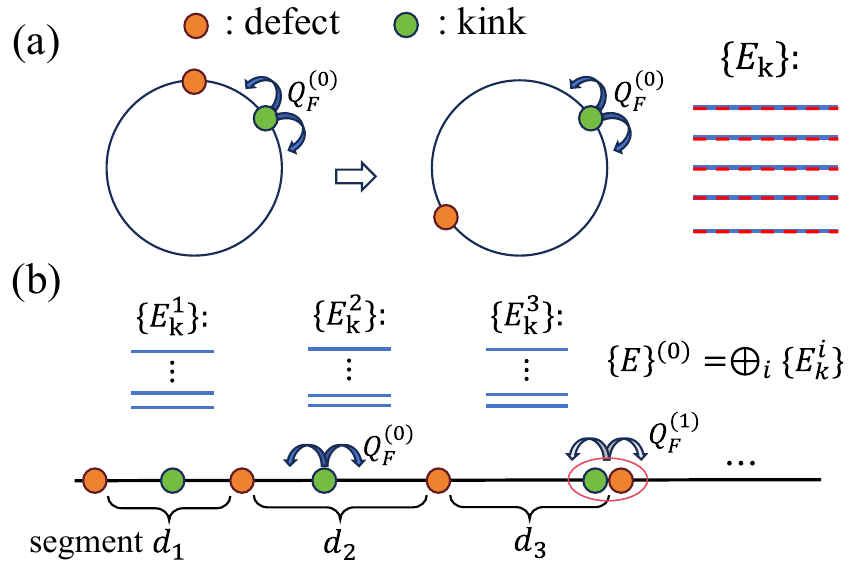}
    \caption{Schematic diagram of the degeneracy structure. (a) One defect and one kink exist in a closed chain (PBC). Here, $Q_F^{(0)}$ suggests that the kink hops freely in the system, while the defect acts as a static cut that turns the system into an open chain. 
    The solid blue line and dashed red line show the energy spectrum of the $Q_F^{(0)}$ in the left and right configurations, respectively. The spectrum remains unchanged regardless of the defect location. The arrows show the dynamics of kinks induced by $Q_F^{(0)}$.
    (b) For a state with multiple defects and kinks, the corresponding energy spectrum $\{E\}^{(0)}$ of the $Q_F^{(0)}$ is composed of the single-particle energy spectrum of each kink. The interval between two defects is defined as a segment.
    The perturbation $Q_F^{(1)}$ causes the defect to move according to the dynamics described by the QMM. It changes the length of each segment $\{d_i\}$, thereby modifying the energy spectrum of the kinks on the segments and consequently changing $\{E\}^{(0)}$.
    }
    \label{fig:degen}
\end{figure}

To illustrate these findings, we first focus on the high-frequency limit $T{\to}0$, where the system is dominated by the lowest order effective Hamiltonian $Q_F^{(0)}$. For simplicity, we begin with an initial state in the sector $\text{GS}_1$ with $\{g_j\}=\{-3, 1, 1,\cdots, 1\}$, in which one defect and one kink co-exist in the system with PBC, as shown in Fig.~\ref{fig:degen} (a). Similar states have also been used in Fig.~\ref{fig:floq} and Fig.~\ref{fig:scaling}. Since the defect is completely frozen, it acts as a static cut that turns the system into an open chain, within which the kink can hop freely. 
Importantly, $Q_F^{(0)}$ always leads to the same spectrum for the kink, regardless of the defect location. Therefore, for a system of $L$ sites, the entire system features an $\mathcal{O}(L)$-fold degeneracy. For any small but finite value of the driving period $T$, those degenerate spectra can now be coupled via terms in $Q_F^{(1)}$, and the defect becomes mobile. The defect thus spreads in space, and degenerate perturbation theory (Sec.~\ref{sec:pert}) predicts that the defect dynamics precisely result in the scaling $\Delta n_j^{\text{d}}\sim(T^2_Ft)^2$, in accordance with the numerical observation in Sec.~\ref{sec.QMMnumerics}. 

Importantly, such a coupling induced by $Q_F^{(1)}$ can be notably suppressed when these degeneracies are lifted. In fact, this naturally happens when the system has multiple defects and kinks, as shown in Fig.~\ref{fig:degen} (b). For a sufficiently large system of length $L$, a finite but low defect density can exist in the initial state. These defects can randomly distribute in space and cut the 1D chain into multiple segments of different lengths $\{d_i\}$. Due to the {local gauge constraint}, one segment must contain precisely one kink, c.f. Sec.~\ref{sec:QMM}, and the entire $Q_F^{(0)}$ spectrum $\{E\}^{(0)}$ is a direct sum of all kink spectra in each segment, $\{E\}^{(0)} = \oplus_i\{E_k^i\}$. Clearly, for different spatial arrangements of $\{d_i\}$, $Q_F^{(0)}$ generally leads to the distinct spectrum $\{E\}^{(0)}$ of the entire system, hence lifting the aforementioned degeneracy. Accidental degeneracy can appear with low probabilities, see discussions in Sec.~\ref{sec:pert}.

To verify the suppressed dynamics when degeneracies are lifted, we consider a scenario where two defects coexist, and the system is initialized in the sector $\text{GS}_2:\{g_j\} = \{-3,1,1,1,1,-3,1,1,1,1\}$. As shown in Fig.~\ref{fig:floq} (a),
the charge $g_j$ remains locally conserved over a long time scale of $10^4$ drive cycles ($1/KT_F=6.2$, thin blue line). In contrast, for $\text{GS}_1$, where degenerate spectra appear, under the same drive frequency (red line), violation of the local charge conservation already becomes noticeable an order of magnitude earlier in time. This comparison thus exemplifies our previous perturbative argument.
Therefore, increasing the number of defects makes the emergence of degenerate structures progressively more difficult, thereby further suppressing defect dynamics.

\section{Discussion}
\label{sec: discuss}
We have presented a general framework to control the dynamics of the local charges in abelian $U(1)$ lattice gauge theories, and hence to protect their experimental realization against perturbations that violate the local symmetry. 
Our theory relies on a Floquet-engineered hierarchical symmetry structure, $U(1)^\text{local} \to \mathbb{Z}^\text{local}_2 \times U(1)^\text{global}$. The 
sub-symmetry induces a set of approximate dynamical selection rules in the local charge dynamics that strongly restrict the inter-sector couplings.  Crucially, the selection rules lead to a sharp contrast in leakage rates out of the different sectors. This reveals an inherent, symmetry-controlled hierarchy in the robustness of LGT {simulations against symmetry-violating terms}.
Also, quantum simulators, which allow for flexible initial-state preparation in distinct sectors and monitoring the real-time dynamics therein, are natural testbeds of our theory. 

As most of these discussions only depend on the local symmetry structure, they can be generalized to systems with large internal d.o.f.~and in higher dimensions, as well as LGTs with classical d.o.f.~\cite{ebner2024eigenstate,homeier2025prethermal}. For instance, in Sec.~\ref{sec: spin-S}, we have numerically verified the sector-dependent robustness of LGTs in systems with spin-1 d.o.f. on the gauge sites. Further, it would be of particular interest to generalize our discussion to non-Abelian $SU(N)$ LGTs~\cite{banerjee2013atomic}, where the symmetry hierarchy can be much richer than the current setting. Exploring the selection rules, together with the resulting constrained dynamics of the local charges, in $SU(N)$ LGTs represents an intriguing avenue for future research.

The selection rules and the resulting constrained charge dynamics
have a wide range of useful implications across different research fields: 

(i) {Local charges in certain sectors are completely frozen. In other sectors, they can} become mobile, but their motion is kinetically constrained. As captured by the effective quantum marble model (QMM), we propose that such kinetic constraints arise from inter-sector defect dynamics being only permitted through collisions with intra-sector kink degrees of freedom.

Although many kinetically constrained models have been studied in recent years~\cite{hudomal2020quantum,lin2020quantum, sala2020ergodicity}, these constraints are typically imposed directly on the underlying physical degrees of freedom, for example, the Rydberg excitations in PXP models. In sharp contrast, here the local charges $G_j$ are emergent degrees of freedom, i.e., composite objects formed from both matter and gauge fields. In general, it remains largely unknown when and how kinetic constraints can arise for emergent degrees of freedom, and the QMM represents a crucial step forward toward a more systematic investigation of this question in the near future.

(ii) Transport is an everlasting research topic in condensed matter physics. One direct consequence of the symmetry-controlled hierarchy is its strong impact on the transport properties of charges. As shown in Fig.~\ref{fig:floq} as well as Sec.~\ref{sec: spin-S}, depending on the choice of the initial sector, the defect exhibits either localized or ballistic dynamics. We note that the present numerical study is limited to small system sizes; it would thus be worthwhile to employ semiclassical approximations or TEBD simulations to further explore transport in larger systems with a finite density of defects. From the analytical perspective, establishing a general hydrodynamic description~\cite{gopalakrishnan2018hydrodynamics, Finite-temperaturetransport2021, wienand2024emergence} and elucidating its dependence on different sectors also remains an interesting open question.

(iii) 
Our protection scheme is particularly relevant for quantum-simulation experiments emulating high-energy physics. As shown in Sec.~\ref{sec:physical sector}, our protocol substantially enhances the robustness of the sector where Gauss' law is obeyed. 

(iv) Our work also sheds light on localization and non-equilibrium phases of matter.
Under a perfect local symmetry, a coherent superposition of different sectors can give rise to disorder-free localization~\cite{smith2017disorder, papaefstathiou2020disorder, mcclarty2020disorder, karpov2021disorder, halimeh2022enhancing}. Once the lattice gauge theories are perturbed, disorder-free localization generally becomes a transient prethermal phenomenon with a finite lifetime. The probability distribution over different sectors controls both the transient localization behavior~\cite{chakraborty2022disorder} and the inter-sector coupling as discussed above. The interplay between these effects can lead to rich and nontrivial scenarios of ergodicity breaking and restoration, which would be worthwhile to investigate further.

\textit{Acknowledgments.---}
This work is supported by Quantum Science and Technology-National Science and Technology Major Project
(No. 2024ZD0301800) and by the National Natural Science
Foundation of China (Grant No. 12474214), and by “The Fundamental Research Funds for the Central Universities, Peking University”, and by ``High-performance Computing Platform of Peking University". 
MB and RM were supported in part by grant no. NSF PHY-2309135 to the Kavli Institute for Theoretical Physics (KITP).
MB was funded by the European Union (ERC, QuSimCtrl, 101113633). 
Views and opinions expressed are however those of the authors only and do not necessarily reflect those of the European Union or the European Research Council Executive Agency. Neither the European Union nor the granting authority can be held responsible for them.

\bibliography{Reference}

\clearpage
\newpage

 \let\addcontentsline\oldaddcontentsline
	\cleardoublepage
	\onecolumngrid
 \begin{center}
\textbf{\large{\textit{Supplementary Material} \\ \smallskip
	Protecting Quantum Simulations of Lattice Gauge Theories \\ 
through Engineered Emergent Hierarchical Symmetries}}\\
		\hfill \break
		\smallskip
	\end{center}
	\renewcommand{\thefigure}{S\arabic{figure}}
        \setcounter{figure}{0}
    \renewcommand{\thesection}{SM\;\arabic{section}}
	\setcounter{section}{0}
	\renewcommand{\theequation}{S.\arabic{equation}}
        \setcounter{equation}{0}
    \renewcommand{\thesubsection}{\arabic{subsection}}
	\setcounter{section}{0}
    \tableofcontents
    \setcounter{page}{1}

\section{Lattice gauge theory represented as a spin system}

The traditional fermionic lattice gauge theory~\cite{pichler2016real} is described by $H_{\text{LGT}} = \sum_j\psi_j^{\dagger}U_{j,j+1}\psi_{j+1}+h.c.$ Here $\psi_j$ ($\psi_j^{\dagger}$) are fermionic annihilation (creation) operators on lattice site $j$, and $U_{j,j+1}$ is the operator conjugate to the electric field $E_{j,j+1}$, $[E_{j,j+1}, U_{j,j+1}] = U_{j,j+1}$, defined on the links. The generator of gauge transformations is $G_j^f = E_{j,j+1}-E_{j-1,j}-\psi_j^{\dagger}\psi_j+\frac{1-(-1)^j}{2}$. Here, for simplicity, we don't include the fermion mass and electric field terms. However, with the hierarchical symmetry structure, the selection rules and the picture of effective dynamics we discuss below will still work with the presence of these terms.

To realize the model with quantum simulators, we
truncate the local Hilbert space dimension on the links, and introduce spin operators $\tau_{j,j+1}$, and map $E_{j,j+1}\to\tau_{j,j+1}^z,\; U_{j,j+1}\to\tau_{j,j+1}^+$. We also apply a Jordan-Wigner transformation to trade the matter fermion for a spin-${1}/{2}$ degree of freedom on the sites, $\psi_{j}^\dagger\to\sigma^+$. This leads to the Hamiltonian from Eq.~\eqref{eq:H_LMG} in the main text:

\begin{equation}
\label{eq:sm1}
H_{\text{LGT}} = \sum_j \sigma_j^+\tau_{j,j+1}^+\sigma_{j+1}^-+h.c.
\end{equation}

More generally, one can use spin-$S$ gauge field with the same Hamiltonian in Eq.~\eqref{eq:sm1}, where the operator $\tau_{j,j+1}^{+(-)}$ now becomes the raising (lowering) operator of spin-$S$.
See also discussions regarding our protection scheme in~\ref{sec: spin-S}.

\section{Gauge violation dynamics in the physical sector}
\label{sec:physical sector}

In the main text, we discuss how the emergent local $\mathbb{Z}_2$ symmetry suppresses the dynamics of gauge violation, and we show that for different  sectors, the leakage rates can be different. We have presented a few examples in the main text and here we further study the so-called physical sector, which attracts intensive research interest in the field of the quantum simulation of particle physics.

In quantum simulation of high-energy physics, the state in the physical sector $\text{GS}_p$ is defined as $G_j^f|\psi\rangle=0$, where $G_j^f$ is defined in previous section and $f$ refers to the original fermionic LGT.
With the definition of gauge generator of spin system in the main text $G_j = \tau_{j,j+1}^z-\tau_{j-1,j}^z-\sigma_j^z$ and its eigenvalue $g_j$, the physical sector corresponds to the sector with $\text{GS}_p:\{g_j\} = \{1,-1,1,-1,\cdots\}$.

\begin{figure}[t]
    \centering
    \includegraphics[width=0.7\linewidth]{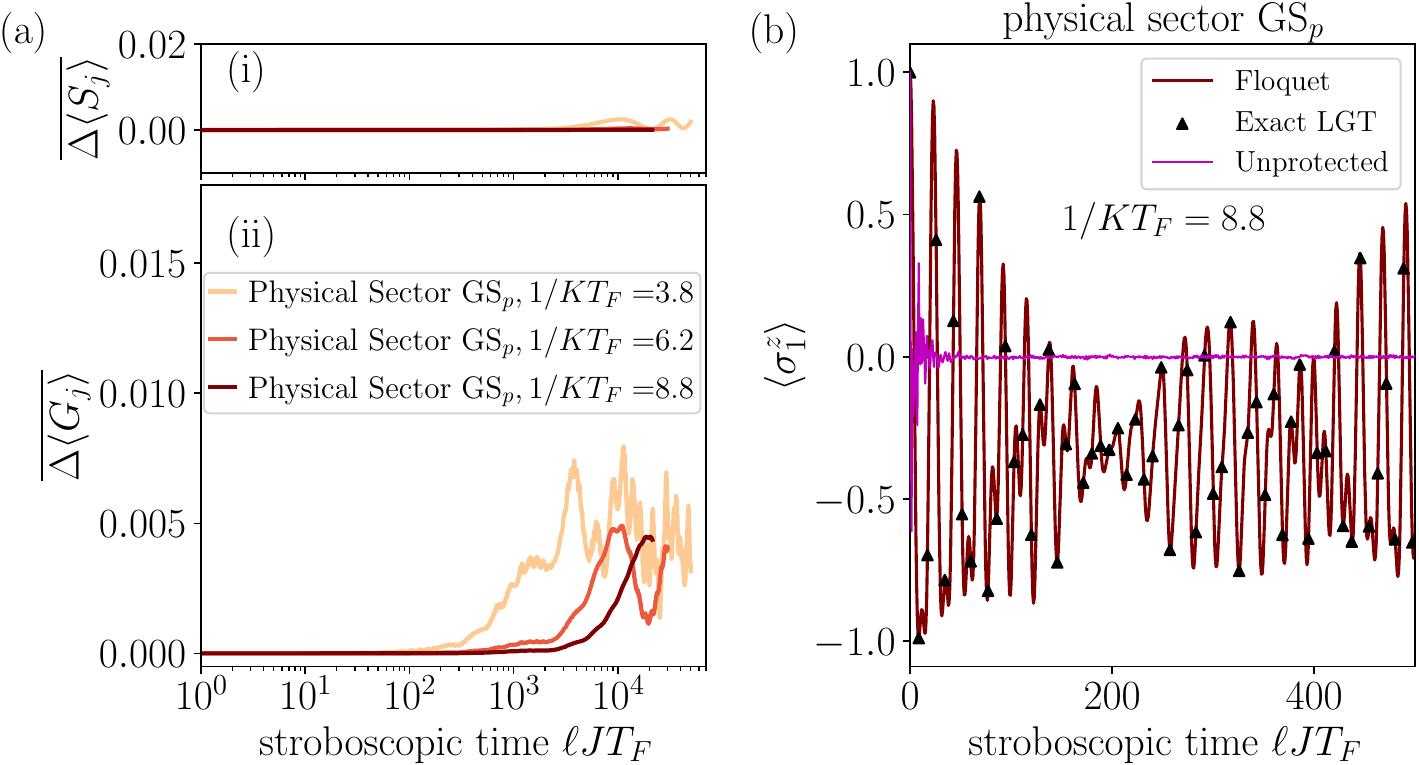}
    \caption{Dynamical protection of matter and gauge field dynamics. (a) Dynamics of gauge violation for the physical sector $\text{GS}_p:\{g_j\}=\{1,-1,1,-1,1,-1,\cdots\}$ with open boundary condition. The inserted figure: dynamics of (i) $\mathbb{Z}_2^\text{local}$ symmetry and (ii) the $U(1)^\text{local}$ symmetry violation for the physical sector.
    The violation is defined as $\overline{\Delta\langle S_j\rangle}=\frac{1}{L}\sum_{j=0}^{L-1}|\langle S_j\rangle-(-1)^{j+1}|$ and $\overline{\Delta\langle G_j\rangle}=\frac{1}{L}\sum_{j=0}^{L-1}|\langle G_j\rangle-(-1)^{j}|$.
    The prethermal lifetime of (i) is much longer than the lifetime of (ii).
Therefore, the dynamics of gauge violation is largely suppressed by the emergent local $\mathbb{Z}_2$ symmetry and the lifetime of the prethermal plateau can be controlled by the driving frequency.
    (b) Comparison between the dynamics of the matter field in the system under the driving protocol in ~\ref{sec:newprotocol} (red lines), the exact LGT (triangle), and the unprotected quench dynamic of $H$ in Eq.~\eqref{eq:lab} (purple line). We can observe the system's dynamics can be well described by LGT under our protocol. However, under the same parameters, the system's quench dynamics without the protocol will rapidly deviate from the LGT dynamics.
    The parameters in the simulation are $J = 1.0$, $K = 4.0,  h = 0.5, \epsilon_1 = \epsilon_2 = 1.0$. System size is $L = 10$.}
    \label{fig:sm-spin1/2}
\end{figure}

In Fig.~\ref{fig:sm-spin1/2}, we show the simulation results of the gauge violation with the system initialized in the physical sector as $\left|_{\Downarrow}\;^{\uparrow}\;_{\Uparrow}\;^{\uparrow}\;_{\Downarrow}\;^{\uparrow}\;_{\Uparrow}\;^{\uparrow}\cdots\right\rangle$. In Fig.~\ref{fig:sm-spin1/2}.(a), the dynamics of gauge violation are significantly suppressed by using our Floquet protection scheme, resulting in a prolonged prethermal plateau even at relatively low driving frequencies. 
In Fig.~\ref{fig:sm-spin1/2}.(b), we demonstrate that the dynamical details of the matter field under our protocol are consistent with those under exact LGT, whereas the simple quench dynamics of the system would cause its dynamics to deviate from exact LGT dynamics rapidly.

\section{selection rules for matter fields and gauge fields} 
In the main text, we discuss the selection rules (Eq.~\eqref{eq.twositesrule}) in the charge dynamics in Sec.~\ref{sec.selectiontrule}.
In this section, we will use spin representations to illustrate these dynamics.

The global $U(1)$ symmetry suggest that the operator $G = \sum_{j}G_j$ is a conserved quantity, and hence the local creation or annihilation of the local charge is not allowed. The local $\mathbb{Z}_2$ symmetry permits the transport of the charge, yet the transport is restricted as $\Delta g {=} 4$. This can be seen by explicitly  
checking the eigenvalue of the local $\mathbb{Z}_2$ operator $S_j = -\tau_{j-1,j}^z\tau_{j,j+1}^z\sigma_j^z$ in the basis state: 
\begin{eqnarray}
    &&g_j = -3, \; \left|^{\uparrow}\;_{\Uparrow}\;^{\downarrow}\right\rangle,\; \text{eigenvalue of }S_j : 1;\nonumber\\
    &&g_j=-1, \;\left|^{\downarrow}\;_{\Uparrow}\;^{\downarrow}\right\rangle,\; 
\left|^{\uparrow}\;_{\Uparrow}\;^{\uparrow}\right\rangle,\;
\left|^{\uparrow}\;_{\Downarrow}\;^{\downarrow}\right\rangle,\; \; \text{eigenvalue of }S_j : -1;\nonumber\\
&&g_j=1, \;\left|^{\downarrow}\;_{\Uparrow}\;^{\uparrow}\right\rangle,\; 
\left|^{\uparrow}\;_{\Downarrow}\;^{\uparrow}\right\rangle,\;
\left|^{\downarrow}\;_{\Downarrow}\;^{\downarrow}\right\rangle,\; \; \text{eigenvalue of }S_j : 1;\nonumber\\
&&g_j=3, \;\left|^{\downarrow}\;_{\Downarrow}\;^{\uparrow}\right\rangle,\; \; \text{eigenvalue of }S_j : -1.
\end{eqnarray}
Clearly, terms that preserve local $\mathbb{Z}_2$ symmetry only connect the sector with $g_j = -3,1$ or $g_j=-1, 3$.

Based on these selection rules, it is natural to understand the cases 1(a) and 1(b) in Sec.~\ref{sec.selectiontrule}, since the condition $\Delta g=4$ cannot be satisfied. 
For the defect with $g_i = -3$, cf.~case (2), charge transfer is allowed, leading to 
$\{g_i\}{=}\{\dots, 1,-3,1,\dots\}\to\{\dots,1, 1,-3,\dots\}$. The corresponding dynamics reads 
\begin{equation}
    | \!\cdot\!\cdot\!\cdot\! \;^{\uparrow_{j-1,j}} \; _{\Uparrow_j} \;^{\downarrow_{j,j+1}} \; _{\Downarrow_{j+1}} \;^{\downarrow_{j+1,j+2}} \!\cdot\!\cdot \cdot\! \rangle \mapsto
    | \!\cdot\!\cdot\!\cdot\! \;^{\uparrow_{j-1,j}} \; _{\Downarrow_j} \;^{\uparrow_{j,j+1}} \; _{\Uparrow_{j+1}} \;^{\downarrow_{j+1,j+2}} \!\cdot\!\cdot \cdot\! \rangle.
\end{equation}
For case 3, we notice that the isolated defect of type $g_j{=}{-}1$ first transfer the charge to the neighboring site with $g_i{=}1$, and the resulting charge configuration becomes $g_j{=}3$ and $g_i{=}{-}3$ through
\begin{eqnarray*}
   | \!\cdot\!\cdot\!\cdot\! \;^{\uparrow_{j-1,j}} \; _{\Uparrow_j} \;^{\uparrow_{j,j+1}} \!\cdot\!\cdot \cdot\! \rangle_{g_j{=}{-}1} 
   &\ \mapsto\ &  
   |\!\cdot\!\cdot\!\cdot\! \;^{\downarrow_{j-1,j}} \; _{\Downarrow_j} \;^{\uparrow_{j,j+1}} \!\cdot\!\cdot \cdot\! \rangle_{g_j{=}3}
   \nonumber \\
   |\!\cdot\!\cdot\!\cdot\! \;^{\downarrow_{j-1,j}} \; _{\Downarrow_j} \;^{\downarrow_{j,j+1}} \!\cdot\!\cdot \cdot\!  \rangle_{g_j{=}1} 
   &\ \mapsto\ &  
   |\!\cdot\!\cdot\!\cdot\! \;^{\uparrow_{j-1,j}} \; _{\Uparrow_j} \;^{\downarrow_{j,j+1}} \!\cdot\!\cdot \cdot\!  \rangle_{g_j{=}{-}3} .
\end{eqnarray*}
Recall that $g_j{=}3$ defects are localized, but defect $g_j{=}{-}3$ can propagate by exchanging charges with neighboring sites.
Therefore, the charge dynamics are dominated by the dynamics of $g_j{=}{-}3$ defects. That said, this charge dynamics can be further suppressed by Floquet engineering, as we show in the main text.

\section{Mapping to quantum marble model}
\label{sec:QMM}
In the main text, we provide the expression for the effective Quantum Marble Model (QMM), which can exactly describe the local $\mathbb{Z}_2$ preserving dynamics in our driving protocol in configurations containing a single defect and kink. Here, we present the explicit mapping from the effective Hamiltonian to the QMM, not only for the case with a single defect and kink in the open chain, but also for the closed chain and for multiple defects and kinks. 

Within the sector with only $g_j=-3, 1$ charges, we have the following local basis
\begin{equation}
    g_j=-3:\; \left|^{\uparrow}\;_{\Uparrow}\;^{\downarrow}\right\rangle;\qquad g_j=1:\;\left|^{\downarrow}\;_{\Uparrow}\;^{\uparrow}\right\rangle,\; \left|^{\uparrow}\;_{\Downarrow}\;^{\uparrow}\right\rangle,\; \left|^{\downarrow}\;_{\Downarrow}\;^{\downarrow}\right\rangle.
    \label{eq:loc_basis}
\end{equation}
Here $\left|^{\downarrow}\;_{\Uparrow}\;^{\uparrow}\right\rangle$ is defined as kink configuration.
Note that since the states here share a gauge-field spin $\tau$, within the sector $\{g_j=-3, 1\}$, a segment, defined as a subsystem with endpoints at the two nearest defects, must contain one kink configuration. This is because, within this sector, the right side of the defect configuration can only be the kink configuration or the configuration $\left|^{\downarrow}\;_{\Uparrow}\;^{\downarrow}\right\rangle$, while the left side of the defect configuration can only be the kink configuration  $\left|^{\downarrow}\;_{\Uparrow}\;^{\uparrow}\right\rangle$ or the configuration $\left|^{\uparrow}\;_{\Uparrow}\;^{\uparrow}\right\rangle$.

\subsection{Open chains with single defect and kink}

We first show that under OBC the Hamiltonians $H_{\text{LGT}}$ and $H_1$ can be mapped to simple hopping models for the kink and defect by defining the creation operators for kink $\kappa_j^{\dagger}$ and defect $\Delta_j^{\dagger}$ as 
\begin{align}
\text{Case 1 (defect on the left-hand side of kink):}\qquad
\kappa_j^{\dagger} &= \prod_{k=j}^{L-1}\tau_{k,k+1}^+\sigma_j^+,\; \Delta_j^{\dagger} = \prod_{k=0}^{j}\tau_{k-1, k}^+\sigma_j^+\nonumber\\
\text{Case 2 (defect on the right-hand side of kink):}\qquad \kappa_j^{\dagger} &= \prod_{k=0}^{j}\tau_{k-1,k}^-\sigma_j^+,\; \Delta_j^{\dagger} = \prod_{k=j}^{L-1}\tau_{k,k+1}^-\sigma_j^+,\nonumber
\end{align}

Let us first focus on Case 1. The Hamiltonians governing the hopping dynamics of \zp{the kink ($\kappa_j^{\dagger}$) and the defect ($\Delta_j^{\dagger}$)} are given by
\begin{eqnarray}
    H_k &=& \sum_{j}\kappa_j^{\dagger}\kappa_{j+1}+h.c = \left(\prod_{k=j+1}^{L-1}\tau_{k,k+1}^+\tau_{k,k+1}^-\right)\sigma_j^+\tau_{j,j+1}^+\sigma_{j+1}^-+h.c.
    \label{eq:Ha}\\
    H_d &=& \sum_j\Delta_j^{\dagger}\Delta_{j+1}+h.c = \left(\prod_{k=0}^j\tau_{k-1, k}^+\tau_{k-1,k}^-\right)\sigma_j^+\tau_{j,j+1}^-\sigma_{j+1}^-+h.c.
    \label{eq:Hb}
\end{eqnarray}
For Case 1, the configuration in the sector $\{g_i\}=\{\dots, 1,-3,1,\dots\}$ is
\begin{equation}
    \left|\;^{\uparrow} \; _{\Downarrow} \;^{\uparrow}\; _{\Downarrow} \;^{\uparrow}\; _{\Downarrow} \;^{\uparrow}\cdots \underbrace{\;^{\uparrow} \; _{\Uparrow} \;^{\downarrow}}_{\text{defect}} \; _{\Downarrow} \;^{\downarrow}\cdots \;^{\downarrow}\; _{\Downarrow} \underbrace{\;^{\downarrow}\; _{\Uparrow} \;^{\uparrow}}_{\text{kink}}\cdots\;^{\uparrow}\;_{\Downarrow} \;^{\uparrow}\; _{\Downarrow} \;^{\uparrow}\; _{\Downarrow} \;^{\uparrow}\right\rangle,
    \label{eq:state}
\end{equation}
from which we see that the gauge field on the left side of the defect and on the right side of the kink are both all spin-up. Hence, the projectors $\left(\prod_{k=j+1}^{L-1}\tau_{k-1,k}^+\tau_{k-1,k}^-\right)$ and $\left(\prod_{k=0}^j\tau_{k-1, k}^+\tau_{k-1,k}^-\right)$ of the gauge field in Eq.~\eqref{eq:Ha} and Eq.~\eqref{eq:Hb} reduce to the identity, leading to the exact relations $H_k=H_{\text{LGT}}$ and $H_k+H_d=H_1$.  

Analogously, for Case 2, the hopping Hamiltonians read as
\begin{eqnarray}
    H_k &=& \sum_j\kappa_j^{\dagger}\kappa_{j+1}+h.c = \left( \prod_{k=0}^j\tau_{k-1,k}^-\tau_{k-1,k}^+ \right)\sigma_j^+\tau_{j,j+1}^+\sigma_{j+1}^-+h.c.
    \label{eq:Ha2}\\
    H_d &=& \sum_jb_j^{\dagger}b_{j+1}+h.c = \left(\prod_{k=j+1}^{L-1}\tau_{k,k+1}^-\tau_{k,k+1}^+\right)\sigma_j^+\tau_{j,j+1}^-\sigma_{j+1}^-+h.c. 
    \label{eq:Hb2}
\end{eqnarray}
The configuration in the sector $\{g_i\}=\{\dots, 1,-3,1,\dots\}$ is 
\begin{equation}
    \left|\;^{\downarrow} \; _{\Downarrow} \;^{\downarrow}\; _{\Downarrow} \;^{\downarrow}\; _{\Downarrow} \;^{\downarrow}\cdots \underbrace{\;^{\downarrow} \; _{\Uparrow} \;^{\uparrow}}_{\text{kink}} \; _{\Downarrow} \;^{\uparrow}\cdots \;^{\uparrow}\; _{\Downarrow} \underbrace{\;^{\uparrow}\; _{\Uparrow} \;^{\downarrow}}_{\text{defect}}\cdots\;^{\downarrow}\;_{\Downarrow} \;^{\downarrow}\; _{\Downarrow} \;^{\downarrow}\; _{\Downarrow} \;^{\downarrow}\right\rangle,
    \label{eq:state2}
\end{equation}
from which we see that the gauge field values on the left side of the defect and on the right side of the kink are all spin-down. Hence, the projectors $\left(\prod_{k=j+1}^{L-1}\tau_{k,k+1}^-\tau_{k,k+1}^+\right)$ and $\left(\prod_{k=0}^j\tau_{k-1, k}^-\tau_{k-1,k}^+\right)$ of the gauge field in Eq.~\eqref{eq:Ha2} and Eq.~\eqref{eq:Hb2} again reduce to the identity, leading to the exact relations $H_k=H_{\text{LGT}}$ and $H_k+H_d=H_1$.

One can further check the mapping for the first-order effective Hamiltonian. The effective Hamiltonian for the Floquet-engineered lattice gauge theory is 
\begin{eqnarray}
    H_{\text{eff}} &=&  Q_F^{(0)}+Q_F^{(1)}, \nonumber\\
    Q_F^{(0)} &=& J\sum_jJ\sigma_j^+\tau_{j,j+1}^+\sigma_{j+1}^-+h.c., \nonumber\\
     Q_F^{(1)} &=& \lambda_0 \sum_j(i\sigma_{j-1}^+\tau_{j-1,j}^-\sigma_j^z\tau_{j,j+1}^+\sigma^-_{j+1}-i\sigma_{j-1}^+\tau_{j-1,j}^+\sigma_j^z\tau_{j,j+1}^-\sigma^-_{j+1}+h.c.),
     \label{eq:simple_effH}
\end{eqnarray}
where $\lambda_0 = \frac{KJ(J+K)}{2(2K+J)}T_F$
. In Case 1, the corresponding first order term is $i\Delta_j^{\dagger}\kappa_{j+1}^{\dagger}\Delta_{j+1}\kappa_{j+2}+h.c.$. Using the mapping we defined before, we obtain
\begin{equation}
\label{eq:case1}
i\Delta_j^{\dagger}\kappa_{j+1}^{\dagger}\Delta_{j+1}\kappa_{j+2}+h.c = i\left(\prod_{k=0}^j\tau_{k-1,k}^+\tau_{k-1, k}^-\right)\sigma_j^+\tau_{j,j+1}^-\sigma_{j+1}^+\sigma_{j+1}^-\tau_{j+1,j+2}^+\sigma_{j+2}^-\left(\prod_{k=j+2}^{L-1}\tau_{k,k+1}^+\tau_{k, k+1}^-\right)+h.c.
\end{equation}
which gives the same results as $Q_F^{(1)}$ when acting on the state in Eq.~\eqref{eq:state} because the second term of $Q_F^{(1)}$ acts as zero on the state. 

In Case 2, the corresponding first order term is $-i\kappa_j^{\dagger}\Delta_{j+1}^{\dagger}\kappa_{j+1}\Delta_{j+2}+h.c$, and we get
\begin{equation}
\label{eq:case2}
-i\kappa_j^{\dagger}\Delta_{j+1}^{\dagger}\kappa_{j+1}\Delta_{j+2}+h.c = -i\left(\prod_{k=0}^j\tau_{k-1,k}^-\tau_{k-1, k}^+\right)\sigma_j^+\tau_{j,j+1}^+\sigma_{j+1}^+\sigma_{j+1}^-\tau_{j+1,j+2}^-\sigma_{j+2}^-\left(\prod_{k=j+2}^{L-1}\tau_{k,k+1}^-\tau_{k, k+1}^+\right)+h.c.
\end{equation}
which gives the same results as $Q_F^{(1)}$ when acting on the state in Eq.~\eqref{eq:state2} because the first term of $Q_F^{(1)}$ acts as zero on the state. 

Therefore, we can use the mapping defined above to effectively describe the original dynamics with the emergent kinks and defects, which can simplify our analysis of the dynamics.

\subsection{Mapping for closed chain and multiple defects and kinks}

From the discussion in the main text, we know that the dynamics of the charges in the system under $\mathbb{Z}_2^{\text{local}}$ symmetry are primarily described by the dynamics of kinks and defects. However, although the mapping of the operators defined in the previous section is direct, it only works for the case with one defect and one kink.
To generalize the effective dynamics above, we can directly map the kink and defect configuration to states as follows:
\begin{equation}
\label{eq:state_map}
    \text{defect: }\left|\;^{\uparrow}\;_{\Uparrow}\;^{\downarrow}\right\rangle\to|2\rangle,\qquad \text{kink: }\left|\;^{\downarrow}\;_{\Uparrow}\;^{\uparrow}\right\rangle\to|1\rangle,\qquad \text{vacuum: }\left|\;^{\downarrow}\;_{\Uparrow}\;^{\downarrow}\right\rangle\to|0\rangle,\;\left|\;^{\uparrow}\;_{\Uparrow}\;^{\uparrow}\right\rangle\to|\tilde{0}\rangle.
\end{equation}
Therefore, in this basis, we can represent the creation operator of defects ($\Delta_j^{\dagger}$) and kinks ($\kappa_j^{\dagger}$) as 
\begin{equation}
    \kappa_j^{\dagger} = |1\rangle_j\langle0|_j+|1\rangle_j\langle\tilde{0}|_j,
    \qquad
    \Delta_j^{\dagger} = |2\rangle_j\langle0|_j+|2\rangle_j\langle\tilde{0}|_j,
\end{equation}
which indicates that $\kappa_j^{\dagger}\Delta_j^{\dagger} = \Delta_j^{\dagger}\kappa_j^{\dagger} = 0$. 

Notice that since the states in Eq.~\eqref{eq:state_map} share a gauge-field spin $\tau$, the arrangement of the mapped states must satisfy certain constraints: 
\begin{itemize}
    \item[(1)] between every nearest pair of $|2\rangle_j$, there must be one $|1\rangle_i$ and vice versa;
    \item[(2)] $|0\rangle_j$ ($|\tilde{0}\rangle_j$) appears only to the right (left) of $|2\rangle_i$ and to the left(right) of $|1\rangle_k$.
\end{itemize}
To implement the constraint (2), we define a projector $P$ to make sure that, between each pair of defect and kink, the vacuum state is properly chosen,
\begin{equation}
    P(|\tilde{0}\;2\;0\rangle+|0\;2\;0\rangle) = |\tilde{0}\;2\;0\rangle,\qquad P(|0\;1\;\tilde{0}\rangle+|\tilde{0}\;1\;0\rangle) = |0\;1\;\tilde{0}\rangle.
\end{equation}
Therefore, only when the dynamics induced by the mapped Hamiltonian preserve these constraints is such a map in Eq.~\eqref{eq:state_map} well-defined. We will demonstrate that the action of the mapped Hamiltonian is identical to that of the original Hamiltonian in the sector $\{g_j\}, g_j=-3,1$, and that the mapped Hamiltonian preserves the aforementioned constraints.

Consider a state with a kink at site $j$ and vacuum states at sites $j-1$ and $j+1$. We have the mapping
\begin{equation}
    \left|\cdots\;^{\downarrow}\underset{j-1}{\;_{\Downarrow}}
    \;^{\downarrow}\underset{j}{\;_{\Uparrow}}\;^{\uparrow}
    \underset{j+1}{\;_{\Downarrow}}\;^{\uparrow}\cdots
    \right\rangle\;\to\;
    \left|\cdots 0\;1\;\tilde{0}\cdots\right\rangle
\end{equation} 
It is straightforward to verify that $H_{\text{LGT}} = \sum_i\sigma_i^+\tau_{i,i+1}^+\sigma_i^-+h.c.$ and the projected $PH_kP = P(\sum_{i}\kappa_i^{\dagger}\kappa_i+h.c.)P$ give the same result when they act on the states:
\begin{eqnarray}
    H_{\text{LGT}}\left|\cdots\;^{\downarrow}\underset{j-1}{\;_{\Downarrow}}
    \;^{\downarrow}\underset{j}{\;_{\Uparrow}}\;^{\uparrow}
    \underset{j+1}{\;_{\Downarrow}}\;^{\uparrow}\cdots
    \right\rangle 
    &=& 
    \left|\cdots\;^{\downarrow}\underset{j-1}{\;_{\Downarrow}}
    \;^{\downarrow}\underset{j}{\;_{\Downarrow}}\;^{\downarrow}
    \underset{j+1}{\;_{\Uparrow}}\;^{\uparrow}\cdots
    \right\rangle + \left|\cdots\;^{\downarrow}\underset{j-1}{\;_{\Uparrow}}
    \;^{\uparrow}\underset{j}{\;_{\Downarrow}}\;^{\uparrow}
    \underset{j+1}{\;_{\Downarrow}}\;^{\uparrow}\cdots
    \right\rangle \nonumber\\
    &=& \left|\cdots 1\;\tilde{0}\;\tilde{0}\cdots\right\rangle + \left|\cdots 0\;0\;1\cdots\right\rangle \nonumber\\
    &=& PH_kP\left|\cdots 0\;1\;\tilde{0}\cdots\right\rangle.
\end{eqnarray}
Similarly, we can also verify that $H_{1}' = \sum_i\sigma_i^+\tau_{i,i+1}^-\sigma_i^-+h.c.$ is equivalent to $PH_dP = P(\sum_{i}\Delta_i^{\dagger}\Delta_i+h.c.)P$.

Furthermore, if we only care about the dynamics of kinks and defects (with spin-up matter field), since the Hamiltonians $H_k$ and $H_d$ are independent of the configurations of the vacuum states $|0\rangle, |\tilde{0}\rangle$, we can simply label the two vacuum states $|0\rangle, |\tilde{0}\rangle$ as the same (degenerate) state $|0\rangle$ and redefine the creation operator of defects ($\Delta_j^{\dagger}$) and kinks ($\kappa_j^{\dagger}$) as 
\begin{equation}
\label{eq:new_mapping}
    \kappa_j^{\dagger} = |1\rangle_j\langle0|_j,
    \qquad
    \Delta_j^{\dagger} = |2\rangle_j\langle0|_j.
\end{equation}
Then we can simply use $H_k$ and $H_d$ to effectively describe the dynamics.

\subsection{Numerical results for multiple defects and kinks dynamics}
\label{sec:multi_def}
\begin{figure*}[t]
    \centering
    \includegraphics[width=0.9\linewidth]{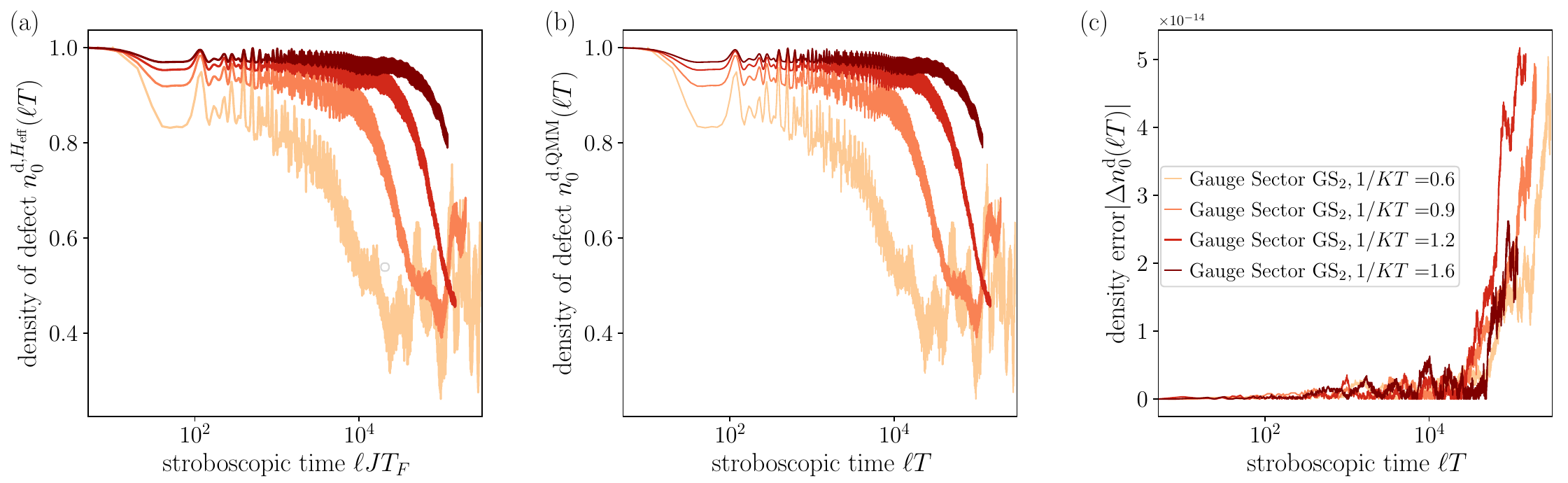}
    \caption{Comparison of the dynamics of the effective Hamiltonian and the QMM. (a)  Defect density at site 0 $n_0^{\text{d}, H_{\text{eff}}} = (1-\langle G_0\rangle)/4$, evolved by the effective Hamiltonian defined in Eq.~\eqref{eq:simple_effH}. (b) Defect density at site 0 $n_0^{\text{d, QMM}} = \langle b_0^{\dagger}b_0\rangle$, evolved by the  corresponding QMM with the mapping define in Eq.~\eqref{eq:new_mapping}. (c) Difference between these two results, $|\Delta n_0^{\text{d}}| = |n_0^{\text{d}, H_{\text{eff}}} - n_0^{\text{d, QMM}}|$. Throughout the entire evolution, their difference is essentially zero, illustrating that QMM precisely describes the dynamics corresponding to the effective Hamiltonian. It further captures the Floquet dynamics when higher-order perturbations are negligible.
    The parameters in the simulation are $J = 1.0$, $K = 4.0$. System size is $L = 10$.
    }
    \label{fig:Heff_QMM_comp}
\end{figure*}

For systems with multiple defects and kinks, with exact mapping as above, our numerical results demonstrate that the QMM can still describe the early dynamics of such systems. Specifically, we consider $h=0$, and compare the dynamics of the effective Hamiltonian in Eq.~\eqref{eq:simple_effH} with the corresponding QMM by simulating the system with the driving protocol in Eq.~\eqref{eq:dirve_simple}. We initialized our system in the sector $\text{GS}_2:\{g_j\} = \{-3, 1, 1,1,1,-3,1,1,1,1\}$ and show the dynamics of the effective Hamiltonian, the QMM and their difference in Fig.~\ref{fig:Heff_QMM_comp}, respectively. The results in the figure demonstrate that the QMM indeed describes the dynamics of the system induced by the effective Hamiltonian throughout the entire evolution, even in the presence of multiple defects and kinks.
So that we can convert the many-body dynamics effectively into a few-body dynamics composed of a small number of defects and kinks configurations, which can facilitate our understanding of the dynamics and enable the large-scale numerical simulations.

\section{Floquet protocol for systems with general perturbation ($h\neq0$)}

\label{sec:newprotocol}
In the main text, we modify the driving protocol to echo out the potentially existing local $\mathbb{Z}_2$-violating terms and achieve an approximate hierarchical symmetry breaking structure: $U(1)^\text{local} \to \mathbb{Z}^\text{local}_2 \times U(1)^\text{global}$, so that the kinetic constraints will still present. In the following section, we show the details of the driving protocol and the effective Hamiltonian. 
\subsection{Driving protocol}
The explicit form of our driving protocol is
\begin{eqnarray}
    &&U_0 = e^{-iH\frac{J+K}{K}T},\nonumber\\
    &&U_1 = P_{\tau}^{z\dagger}P_{\tau}^{x\dagger}e^{-iHT}P_{\tau}^{x}P_{\tau}^{z}\equiv e^{-iH_{(1)}T},\nonumber\\
    &&U_2 = P_{\tau}^{z\dagger}P_{\sigma}^{z\dagger}e^{-iH\frac{J}{K}T}P_{\sigma}^{z}P_{\tau}^{z}\equiv e^{-iH_{(2)}\frac{J+K}{K}T}, \\
    &&U_3 = P_{\tau}^{x\dagger}P_{\sigma}^{z\dagger}e^{-iHT}P_{\sigma}^{z}P_{\tau}^{x} \equiv e^{-iH_{(3)}T},\nonumber\\
    &&U_F = U_0U_1U_2U_3U_2U_3U_0U_1 \equiv e^{-iQ_FT_F},\nonumber
\end{eqnarray}
where $H = JH_{\text{LGT}}+KH_1+hH_0$,  $H_{\text{LGT}} = \sum_j\sigma_j^+\tau_{j,j+1}^+\sigma_{j+1}^-+h.c.$ preserves local $U(1)$ symmetry, $H_1 = \sum_j\sigma_j^+\tau_{j,j+1}^x\tau_{j+1}^-+h.c.$ preserves local $\mathbb{Z}_2$ symmetry and global $U(1)$ symmetry, and 
$H_0 = \sum_{j}\tau_{j,j+1}^x+\epsilon_1\sigma_j^z\tau_{j,j+1}^x+\epsilon_2\tau_{j,j+1}^x\sigma_{j+1}^z$ further breaks these symmetries. The driving period is
$T_F = 8(2+\frac{J}{K})T$. The pulses we use in the driving protocol are
\begin{equation}
    P_{\tau}^z = \exp(-i\frac{\pi}{2}\sum_j\tau_{j,j+1}^z), \quad P_{\tau}^x = \exp(-i\frac{\pi}{2}\sum_j\tau_{j,j+1}^x), \quad P_{\sigma}^z = \exp(-i\frac{\pi}{2}\sum_{j\in \text{odd sites}}\sigma_{j}^z),
\end{equation}
where $\tau_{j,j+1}^{x,z}$, $\sigma_j^z$ are Pauli matrices representing the degree of freedom of the gauge and matter fields. The effective Hamiltonians $H_{(i)}$ for each sequence $U_{i}$ in the driving protocol are
\begin{eqnarray}
    &&H_{(1)} = JH_{\text{LGT}} - (J+K)H_1 - H_0,\nonumber\\
    &&H_{(2)} = JH_{\text{LGT}} + KH_1 - H_0,\\
    &&H_{(3)} = JH_{\text{LGT}} - (J+K)H_1 + H_0.\nonumber
\end{eqnarray}

The effective Hamiltonian for one driving period satisfies
\begin{eqnarray}
    Q_F^{(0)} &=& JH_{\text{LGT}},\nonumber\\
    Q_F^{(1)} &=& 0,\\
    Q_F^{(2)} &=& T_F^2\left(\alpha_1\left[H_{\text{LGT}},\left[H_{\text{LGT}}, H_1\right]\right]+
    \alpha_2\left[H_1,\left[H_{\text{LGT}}, H_1\right]\right]\right.\nonumber\\    &&\left.\;\;\;\;\;\;+\beta_1\left[H_0,\left[H_{\text{LGT}}, H_0\right]\right]+\beta_2\left[H_0,\left[H_1, H_0\right]\right]
    \right)\nonumber,
    \label{eq:eff_Q}
\end{eqnarray}
which is obtained by the high-frequency expansion of $U_F$. The explicit form of $\alpha_{1,2}$ and $\beta_{1,2}$ is shown in Eq.~\eqref{eq: H_eff_SM}. Since there are only quadratic terms of $H_0$ in $Q_F^{(2)}$, the local $\mathbb{Z}_2$ breaking terms are parametrically small 
$\mathcal{O}(h^2/JK)$ (for simplicity, here we assume $\epsilon_1 = \epsilon_2 = 1.0$). Therefore, the effective Hamiltonian $Q_F$ has the approximate local $U(1)^\text{local} \to \mathbb{Z}^\text{local}_2 \times U(1)^\text{global}\to U(1)^\text{global}$ hierarchical symmetry structure in its perturbative expansion. Details of the derivation of the effective Hamiltonian expansion can be found in the next subsection.

In the high-frequency regime, the system enters prethermal plateaus at different time scales with the corresponding emergent symmetries in the hierarchical symmetry structure, respectively. 
This driving protocol and the approximate hierarchical symmetry structure can be generalized to systems with larger spin-$S$ gauge and matter degrees of freedom.

\subsection{Effective Hamiltonian for the driving protocol}
\label{sec: sm4_2}
Using the Baker-Campbell-Hausdorff lemma, one can check that the effective Hamiltonian $Q_F$ has the structure
\begin{eqnarray}
    Q_F &=& Q_F^{(0)}+Q_F^{(1)}+\cdots, \; Q_F^{(n)}=\mathcal{O}(T_F^n)\nonumber\\
    Q_F^{(0)} &=& JH_{\text{LGT}}, \\
    Q_F^{(1)} &=& 0,\nonumber\\
    Q_F^{(2)} &=& \left(-\frac{(J+K)J^2KT_F^2}{128(J+2K)}+\frac{(J+K)KJ^3T_F^2}{768(J+2K)^2}\right)[H_{\text{LGT}}, [H_{\text{LGT}}, H_1]] + \frac{2(J+K)^2K^2JT_F^2}{768(J+2K)^2}[H_1,[H_{\text{LGT}}, H_1]]\nonumber\\
    &&+\frac{h^2(J+K)^2K^2JT_F^2}{384(J+2K)^2}\left(\frac{1}{K(J+K)}+\left(\frac{1}{K+J}-\frac{1}{K}\right)^2\right)[H_0, [H_{\text{LGT}}, H_0]] \nonumber\\
    &&-\frac{h^2(J+K)KJT_F^2}{768(J+2K)^2}[H_0, [H_1, H_0]],\nonumber
    \label{eq: H_eff_SM}
\end{eqnarray}
which gives the explicit form of $\alpha_1,\alpha_2,\beta_1,\beta_2$ in Eq.~\eqref{eq:eff_Q}. 

Here the first two terms serves as $Q_F^{'(2)}$
\begin{equation}
    Q_F'^{(2)} = \left(-\frac{(J+K)J^2KT_F^2}{128(J+2K)}+\frac{(J+K)KJ^3T_F^2}{768(J+2K)^2}\right)[H_{\text{LGT}}, [H_{\text{LGT}}, H_1]] + \frac{2(J+K)^2K^2JT_F^2}{768(J+2K)^2}[H_1,[H_{\text{LGT}}, H_1]],
\end{equation}
which preserves $\mathbb{Z}_2^{\text{local}}$ symmetry. 

We focus on the local $\mathbb{Z}_2$ preserving part of the second order effective Hamiltonian in Eq.~\eqref{eq:eff_Q} $Q_F^{(2)'} = T_F^2(\alpha_1[H_{\text{LGT}}, [H_{\text{LGT}}, H_1]]+\alpha_2[H_1, [H_{\text{LGT}}, H_1]])$. The corresponding QMM is
\begin{eqnarray}
    &&H_{\text{QMM}} = J\sum_{j}(\kappa^{\dagger}_{j+1}\kappa_j+h.c.)+\lambda_1D-\lambda_2D(\kappa^{(\dagger)}\leftrightarrow \Delta^{(\dagger)}),\\
    &&D = \sum_j\left(-2\kappa_j\kappa_{j+1}^{\dagger}\Delta_{j+1}\Delta_{j+2}^{\dagger}\kappa_{j+2}\kappa_{j+3}^{\dagger}-\Delta_j\Delta_{j+1}^{\dagger}\kappa_{j+1}\kappa_{j+3}^{\dagger}-\kappa_{j-1}\kappa_{j+1}^{\dagger}\Delta_{j+1}\Delta_{j+2}^{\dagger}+\Delta_j^{\dagger}\Delta_{j+1}n_{j+2}^k+n_j^k\Delta_{j+1}^{\dagger}\Delta_{j+2}+h.c.\right)\nonumber,
    \label{eq:QMM_EM}
\end{eqnarray}
where $\lambda_1 = T_F^2(\alpha_1+\alpha_2),\;\lambda_2 = T_F^2\alpha_2,\;n_j^{\text{k}} = \kappa_j^{\dagger}\kappa_j,\;n_j^{\text{d}} = \Delta_j^{\dagger}\Delta_j$.
The QMM Hamiltonian effectively describes the early dynamics of our new driving protocol, which includes the presence of further local $\mathbb{Z}_2$-breaking terms. From which we can see that the dynamics of defects highly depend on the dynamics of kinks, which gives rise to emergent kinetic constraints that suppress the spreading of defects.

\section{Perturbation theory for the dynamics of local charge}
\label{sec:pert}
Here, we present a perturbative approach to describe the dynamics of gauge violation within a hierarchical symmetry breaking structure. This is achieved by studying the time evolution of the charge $G_i$ at stroboscopic times $t=\ell T$
\begin{equation}
\label{eq.symmetrycharge}
    G_i(\ell T_F) = (U_F^{\dagger})^{\ell} G_iU_F^{\ell} = (e^{iQ_F^{(0)}T_F}U^{\dagger})^{\ell}G_i(Ue^{-iQ_F^{(0)}T_F})^{\ell},
\end{equation}
where we define $U = U_Fe^{iQ_F^{(0)}T_F}$, which can be perturbatively expanded as 
\begin{equation}
    U\equiv e^{-iQT_F} = 1-iT_F(Q^{(1)}+Q^{(2)}+\cdots) + \frac{(-iT_F)^2}{2!}(Q^{(1)}+Q^{(2)}+\cdots)(Q^{(1)}+Q^{(2)}+\cdots) + \cdots,
\end{equation}
and we used the perturbative expansion $Q = Q^{(0)} + Q^{(1)}+ Q^{(2)}+\cdots,\; Q^{(n)} = \mathcal{O}(T_F^n)$. Note that $Q^{(0)}$ is generally different from $Q_F^{(0)}$.
More concretely, for the driving protocol shown in ~\ref{sec:newprotocol}, one has $Q^{(0)} = 0$, $Q^{(1)} = Q_F^{(1)} = 0$ and $Q^{(2)} = Q_F^{(2)}$, but their higher order terms can differ.
We define $V_n = Q^{(n)}/T_F^n$ to factor out the order of the driving period $T_F$ and define a $T_F$-independent operator. 

Starting from an initial state $\left|\psi_{\text{ini}}\right\rangle$ , the expectation value of the charge evolves as $\left\langle\psi_{\text{ini}}\right| G_i(\ell T_F)\left|\psi_{\text{ini}}\right\rangle$. Since $Q^{(1)} = Q_F^{(1)} = 0$, its evolution is dominated by $V_2 = Q^{(2)}/T^2_F$, and hence we only focus on the contribution of $V_2$, which we indicate as $\langle G_i(\ell T_F)\rangle_{V_2}$.
To obtain the expansion of the drive frequency, we introduce the notation $e^{iQ_F^{(0)}T_F}G_ie^{-iQ_F^{(0)}T_F} = e^{i\mathcal{L}_0T_F}G_i$, where $\mathcal{L}_0[\cdot] = [Q_F^{(0)}, \cdot]$ and $U \approx e^{-iT_F^3V_2},\; U^{\dagger}G_i U = \sum_{m=0}^{\infty}T_F^{3m}\mathcal{A}_{V_2}^mG_i$, where $\mathcal{A}_{V_2}^m[\cdot] = \frac{1}{m!}(i\mathcal{L}_{V_2})^m[\cdot]$ and $\mathcal{L}_{V_2}[\cdot] = [V_2, \cdot]$.
Then we can rewrite the dynamics of the charge $G_i$ in the Heisenberg picture, Eq.~\eqref{eq.symmetrycharge}, as 
\begin{equation}
\label{eq:expansion1}
    G_i(\ell T_F)=\left( e^{i\mathcal{L}_0T_F}\sum_{m=0}^{\infty}T_F^{3m}\mathcal{A}_{V_2}^m\right)^{\ell}G_i.
\end{equation}
We define a perturbative expansion as $\langle G_i(\ell T_F)\rangle_{V_2}=\sum_n\langle G_i(\ell T_F)\rangle_{V_2}^{(n)}$ 
where $\langle G_i(\ell T_F)\rangle_{V_2}^{(n)}\sim \mathcal{O}(T_F^{3n})$. 
Therefore, to obtain the $n$-th order contribution $\langle G_i(\ell T_F)\rangle_{V_2}^{(n)}$, we extract a series of terms in the full expansion in Eq.~\eqref{eq:expansion1} with the form as
\begin{equation}
    \prod_p\left( e^{i\mathcal{L}_0T_F}T_F^{3a_p}\mathcal{A}_{V_2}^{a_p}\right)G_i = T_F^{3\sum_pa_p}\prod_p\left( e^{i\mathcal{L}_0T_F}\mathcal{A}_{V_2}^{a_p}\right)G_i,
\end{equation}
where $a_p\in \mathbb{N}$. Notice that when $a_p = 0$, $\mathcal{A}_{V_2}^{a_p}\equiv \mathcal{I}$. Then if $a_{p_1}, a_{p_2}\ne0$, while $a_{p}=0$ for all $p_1<p<p_2$, we can group the remaining terms into $e^{i\mathcal{L}_0T_Fl_p}, \;l_p = p_2-p_1$. Therefore, we can specify that \(a_p > 0\) and incorporate the case where $a_p = 0$ into the discussion of $l_p$.
For the $n$-th order dynamics of the charge, we have $\sum_{p}a_p=n$ and we need to sum over all the possible sets of $\{a_p\}$ and $\{l_p\}$ to get the final results.
Then the $n$-th order dynamics of the charge are
\begin{equation}
    \left\langle G_i(\ell T_F)\right\rangle^{(n)}_{V_2} = T_F^{3n}\left\langle\psi_{\text{ini}}\right|\sum_{\substack{\{a_p|\sum_pa_p = n\}\\\{l_p|l_p\ge 1,\;\sum_pl_p\le\ell\}\\} }\left(\prod_pe^{i\mathcal{L}_0T_Fl_p}\mathcal{A}_{V_2}^{a_p}\right)e^{i\mathcal{L}_0T_F(\ell-\sum_{p}l_p)}G_i\left|\psi_{\text{ini}}\right\rangle.
    \label{eq:dyna_sm}
\end{equation}

\begin{figure}
    \centering
    \includegraphics[width=0.7\linewidth]{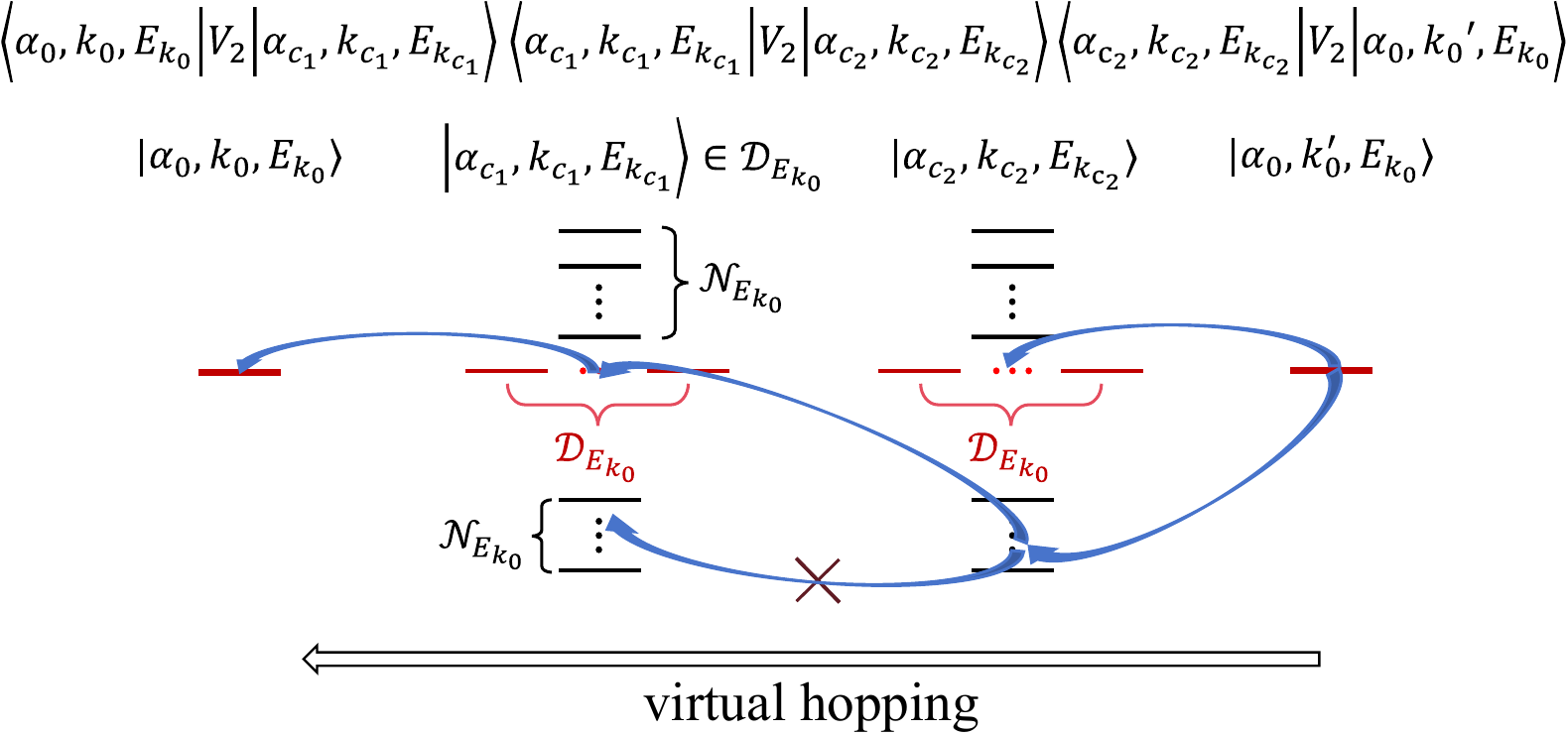}
    \caption{Schematic diagram of the degenerate perturbation theory. Here, as an example, we show the dynamic channel with two virtual processes in high-order expansion, which can be generalized to cases with more intermediate states.
    When we calculate the $n-$th order expansion of Eq.~\eqref{eq:dyna_pert}, we insert the complete basis expanded by  eigenstates $|\alpha_i,k_i,E_{k_i}\rangle$ of $Q_F^{(0)}$; some of these states are chosen to be degenerate with the initial state (red states)  $|\alpha_m,k_m,E_{k_m}\rangle\in \mathcal{D}_{E_{k_0}}, \; E_{k_m} = E_{k_0}$ to get rid of the phase factor in Eq.~\eqref{eq:dyan}, while other
    Non-degenerate eigenstates do not contribute to this dynamical channel.
    The bottom arrow indicates the system's evolution from the initial state to the final state under the perturbation $V_2$. The blue arrows indicate possible dynamical channels evolving from the initial state in the degenerate perturbation expansion. We need to sum over all possible paths to get the final results.
    }
    \label{fig:placeholder}
\end{figure}

We can insert the complete basis, expanded by the eigenstates of $Q_F^{(0)}$ $,\left|\alpha_p, k_p,E_{k_p}\right\rangle$, to simplify this expression. Here $\alpha_p$ is used to label sectors $\text{GS}_{\alpha_p}$ with different charge configurations $\textbf{g} $, $E_{k_p}$ represents the eigenenergy, and $k_p$ refers to the quantum number associated with distinct states at the degenerate energy. 
The following expression appears after we have inserted the complete basis,
\begin{equation}
\label{eq:dyan}
\left\langle\alpha_p, k_p,E_{k_p}\right|
    e^{i\mathcal{L}_0T_Fl}\mathcal{O}
    \left|\alpha_q, k_q, E_{k_q}\right\rangle = e^{iT_Fl(E_{k_p} - E_{k_q})}\left\langle\alpha_p, k_p,E_{k_p}\right|
   \mathcal{O}
    \left|\alpha_q, k_q,E_{k_q}\right\rangle.
\end{equation}
Importantly, according to the Floquet Fermi's Golden rule argument~\cite{ikeda2021fermi, yeh2023decay}, after performing a long-time average, Eq.~\eqref{eq:dyan} becomes proportional to $\delta_F(E_{k_p}-E_{k_q}) = \sum_{j=-\infty}^{\infty}\delta(E_{k_p}-E_{k_q}+{j\pi}/{T_F})$. Hence, when two energy levels fail to match this resonance condition, their coupling will be suppressed. Therefore, in the high-frequency regime, the coupling between two degenerate states naturally dominates the contribution to Eq.~\eqref{eq:dyna_sm}, without absorbing energy from the external high-frequency drive.

For simplicity, we focus on the initial state chosen in one fixed sector $\text{GS}_{\alpha_0}$, and hence the initial state can be decomposed as $\left|\psi_{\text{int}}\right\rangle = \sum_{{k_0}}\lambda_{k_0}\left|\alpha_0, k_0, E_{k_0}\right\rangle$.
For degenerate dynamics, we only count the contribution from the degenerate eigenstates to Eq.~\eqref{eq:dyna_sm} that make the phase factor on the right-hand side of Eq.~\eqref{eq:dyan} vanish. We can then further simplify Eq.~\eqref{eq:dyna_sm}. So that we can not only obtain the scaling of the charge dynamics over time and the lifetime of the prethermal plateau, but also diagnose the relative strengths between different dynamical channels.
\begin{eqnarray}
    &&\left\langle G_i\right\rangle^{(n)}_{V_2}(\ell T_F) =T_F^{3n}\sum_{k_0,k_0'}\lambda_{k_0}^*\lambda_{k_0'}\times\nonumber\\
&&\left[\sum_{m=1}^n\ell^m\left(\sum_{d=m}^nC_{m,d}\sum_{\substack{\{\left|\alpha_j,k_j,E_{k_j}\right\rangle\},\\
\left|\alpha_q,k_q,E_{k_q}\right\rangle\in\mathcal{D}_{E_{k_0}},q \in[1,d-1]}    }
    \mathcal{E}^{(n)}_{\{\left|\alpha_j,k_j,E_{k_j}\right\rangle\}}\left(\left\langle\alpha_0,k_0,E_{k_0}\right|(i\mathcal{L}_{V_2})^nG_i\left|\alpha_0,k_0',E_{k_0}\right\rangle\right)\right)\right],
\label{eq:dyna_pert}
\end{eqnarray}
where $C_{m,d}$ is a constant defined by the expansion in Eq.~\eqref{eq:dyna_sm}, $d$ refers to the number of $\mathcal{A}_{V_2}^{a_p}$ in Eq.~\eqref{eq:dyna_sm} and $\mathcal{D}_{E_{k_0}}$ is the set of degenerate states ($E_{k_q} = E_{k_0}$). Here we sum over all the possible set of eigenstates $\{\left|\alpha_j,k_j,E_{k_j}\right\rangle\}$ chosen from the complete basis of eigenstates of $Q_F^{(0)}$ and satisfy that $d-1$ of the eigenstates belong to $\mathcal{D}_{E_{k_0}}$. 
To illustrate the meaning of the function $\mathcal{E}^{(n)}_{\{\left|\alpha_j,k_j,E_{k_j}\right\rangle\}}$ in Eq.~\eqref{eq:dyna_pert}, we will use the following expansion
\begin{equation}
\label{eq.operator}
(i\mathcal{L}_{V_2})^nG_i = \sum_{j=0}^n\beta_jV_2^{n-j}G_iV_2^j,
\end{equation}
where $\{\beta_j\}$ is a constant defined by the expansion.
Then the function $\mathcal{E}^{(n)}_{\{\left|\alpha_j,k_j,E_{k_j}\right\rangle\}}$ is defined as
\begin{eqnarray}
\label{eq:mathcal_E}
&&\mathcal{E}^{(n)}_{\{\left|\alpha_j,k_j,E_{k_j}\right\rangle\}}\left(\left\langle\alpha_0,k_0,E_{k_0}\right|(i\mathcal{L}_{V_2})^nG_i\left|\alpha_0,k_0',E_{k_0}\right\rangle\right) = \sum_{j=0}^n\beta_j\sum_{\{c_j\}\in P_{n-1}}\alpha_{c_{n-j}}^i
\left\langle\alpha_0,k_0,E_{k_0}\right|V_2\left|\alpha_{c_1},k_{c_1},E_{k_{c_1}}\right\rangle\nonumber\\
&&
\langle\alpha_{c_1},k_{c_1},E_{k_{c_1}}|V_2\left|\alpha_{c_2},k_{c_2},E_{k_{c_2}}\right\rangle\cdots
\langle\alpha_{c_{n-1}},k_{c_{n-1}},E_{k_{c_{n-1}}}|V_2\left|\alpha_{0},k_{0}',E_{k_{0}}\right\rangle,
\end{eqnarray}
$\mathcal{E}^{(n)}_{\{\left|\alpha_j,k_j,E_{k_j}\right\rangle\}}$ can be regarded as a series of dynamical channels under a specified set of intermediate states $\{\left|\alpha_j,k_j,E_{k_j}\right\rangle\}$, as shown in Fig.~\ref{fig:placeholder}.
Here, the summation set $P_{n-1}$ is the $(n-1)$-th order permutation group. Since we can insert the eigenstates of the selected set into $\left\langle\alpha_0,k_0,E_{k_0}\right|(i\mathcal{L}_{V_2})^nG_i\left|\alpha_0,k_0',E_{k_0}\right\rangle$ in any order, we need to sum over the set of $c_j$. And because $[G_i, Q_F^{(0)}] = 0$, $G_i\left|\alpha_{c_{n-j}},k_{c_{n-j}},E_{k_{c_{n-j}}}\right\rangle = \alpha_{c_{n-j}}^i\left|\alpha_{c_{n-j}},k_{c_{n-j}},E_{k_{c_{n-j}}}\right\rangle$, where $\alpha_{c_{n-j}}^i$ represents the charge on site i for the state $\left|\alpha_{c_{n-j}},k_{c_{n-j}},E_{k_{c_{n-j}}}\right\rangle$. 

The QMM representation of $V_2 = Q^{(2)}/T_F^2 = Q_F^{(2)}/T_F^2$ in Eq.~\eqref{eq:QMM_EM} reveals that when $n=1$ in the equation Eq.~\eqref{eq:dyna_pert}, it makes no contribution to the dynamics because $\left\langle\alpha_0,k_0,E_{k_0}\right|i[V_2, G_i]\left|\alpha_{0},k_{0}',E_{k_{0}}\right\rangle = 0$.
Therefore, the leading order dynamics is $n=2$. According to Eq.~\eqref{eq:dyna_pert}, its contribution to the dynamics is 
\begin{eqnarray}
    \left\langle G_i\right\rangle^{(2)}_{V_2}(\ell T_F) = T_F^6(c_1 \ell^2+c_2\ell) = c_1 T_F^4(\ell T_F)^2+ c_2T_F^5(\ell T_F).
\end{eqnarray}
where the coefficients $c_1$ and $c_2$ can be obtained from Eq.~\eqref{eq:mathcal_E}
\begin{eqnarray}
\label{eq:parameter}
    &&c_1 = -\sum_{k_0, k_0'}\lambda_{k_0}^*\lambda_{k_0'}\sum_{|\alpha_1, k_1, E_{k_1}\rangle\in \mathcal{D}_{E_{k_0}}}(\alpha_0^i-\alpha_{1}^i)(\langle \alpha_0,k_0, E_{k_0}|V_2|\alpha_1, k_1, E_{k_1}\rangle\langle \alpha_1,k_1, E_{k_1}|V_2|\alpha_0, k_0', E_{k_0}\rangle),\nonumber\\
    &&c_2 = -\sum_{k_0, k_0'}\lambda_{k_0}^*\lambda_{k_0'}\sum_{|\alpha_1, k_1, E_{k_1}\rangle\in \mathcal{N}_{E_{k_0}}}(\alpha_0^i-\alpha_{1}^i)(\langle \alpha_0,k_0, E_{k_0}|V_2|\alpha_1, k_1, E_{k_1}\rangle\langle \alpha_1,k_1, E_{k_1}|V_2|\alpha_0, k_0', E_{k_0}\rangle),
\end{eqnarray}
where $\mathcal{N}_{E_{k_0}}$ is the set of non-degenerated states ($E_{k_q} \ne E_{k_0}$).
Since the first term makes a significant contribution to the dynamics on the timescale of $T_F^{-2}$, while the second term only becomes significant on timescales of $T_F^{-5}$, we can derive the leading-order dynamics as
\begin{equation}
    \left|\left\langle G_i\right\rangle(t) - \left\langle G_i\right\rangle(0)\right| \propto T_F^4t^2,
\end{equation}
which is verified by the numerical results in Fig.~\ref{fig:scaling}.

For simplicity, in our preceding discussion of leading-order dynamics, we did not focus on any particular term within the dynamical channels. However, the specific form of QMM can help us determine whether a particular dynamical channel is suppressed. Furthermore, QMM maps $Q_F^{(0)}$ to a particle hopping problem with defects, which aids in identifying degenerate structures within the system under different defect configurations. This enables us to determine under which conditions the dynamics dominated by degenerate energy levels, distinct from general Floquet dynamics, will prevail.

On longer timescales, higher-order perturbations will influence the dynamics of the system. We provide a higher-order expansion here to offer a concrete example for the long-time evolution of the system and also help readers understand the expansion in Eq.~\ref{eq:dyna_pert}.
\begin{eqnarray}
    &&\left\langle G_i\right\rangle^{(3)}_{V_m}(\ell T_F) =T_F^{3(m+1)}\sum_{k_0,k_0'}\lambda_{k_0}^*\lambda_{k_0'}
\left[\left(
\frac{\ell^3}{6}\sum_{\substack{
\left|\alpha_{1},k_1,E_{k_1}\right\rangle\in\mathcal{D}_{E_{k_0}}\\
\left|\alpha_{2},k_2,E_{k_2}\right\rangle\in\mathcal{D}_{E_{k_0}}}}
+\left(\frac{\ell^2}{4}-\frac{\ell}{6}\right)\sum_{\substack{
\left|\alpha_{1},k_1,E_{k_1}\right\rangle\in\mathcal{D}_{E_{k_0}}\\
\left|\alpha_{2},k_2,E_{k_2}\right\rangle\in\mathcal{N}_{E_{k_0}}}} + \frac{\ell}{6}\sum_{\substack{
\left|\alpha_{1},k_1,E_{k_1}\right\rangle\in\mathcal{N}_{E_{k_0}}\\
\left|\alpha_{2},k_2,E_{k_2}\right\rangle\in\mathcal{N}_{E_{k_0}}}}
\right)\times\right.\nonumber\\
&&\left.\mathcal{E}^{(3)}_{\{\left|\alpha_j,k_j,E_{k_j}\right\rangle\}}\left(\left\langle\alpha_0,k_0,E_{k_0}\right|(i\mathcal{L}_{V_2})^3G_i\left|\alpha_0,k_0',E_{k_0}\right\rangle\right)\right].
\end{eqnarray}

In the main text, we show that the 0-th order effective Hamiltonian $Q_F^{(0)}$ only involves the dynamics of kinks, while higher order effective Hamiltonians introduce the dynamics of defects governed by the QMM Hamiltonian. Therefore, the entire $Q_F^{(0)}$ spectrum is a direct sum of the kink spectrum in each segment, with segment lengths $\{d_i\}$, cf. Fig.~\ref{fig:degen}. When defects move, $\{d_i\}$ will be rearranged and hence the degeneracy is usually lifted.

However, accidental degeneracy can still arise if, e.g., 
1.~all defects move by the same distance in the same direction, leaving $\{d_i\}$ unchanged; 
2.~two neighboring segments interchange their lengths, e.g.,
$\{\cdots, d_{i-1}, d_{i},\cdots\}$ and $\{\cdots, d_{i}, d_{i-1},\cdots\}$. 
These exceptions generally occur with low probability. 
Both cases usually correspond to higher-order dynamics unless the required distance of defect movement is small. However, these degenerate dynamics may still dominate the dynamics of the defects, since other non-degenerated states are still suppressed by the aforementioned delta function $\sum_{p=-\infty}^{\infty}\delta(E_{k_n}-E_{k_m}+{p\pi}/{T_F})$.

For extremely low density of defects, the length of each segment $\{d_i\}$ tends to diverge in the thermodynamic limit, $L\to\infty$; this makes the kink spectrum in each individual segment continuous, and accidental degeneracy can easily appear. Note, this does not necessarily mean that the defect hopping can easily occur. According to Eq.~\eqref{eq.QMMeffective}, defect can move only when it collides with the kink. However, within a single segment, the kink tends to delocalize and hence the probability of having a kink sitting at the edge of a segment decays algebraically for larger segment size. Therefore, it remains an interesting open question to further explore the stability of LGTs in this regime.

\begin{figure*}[t]
    \centering
    \includegraphics[width=\linewidth]{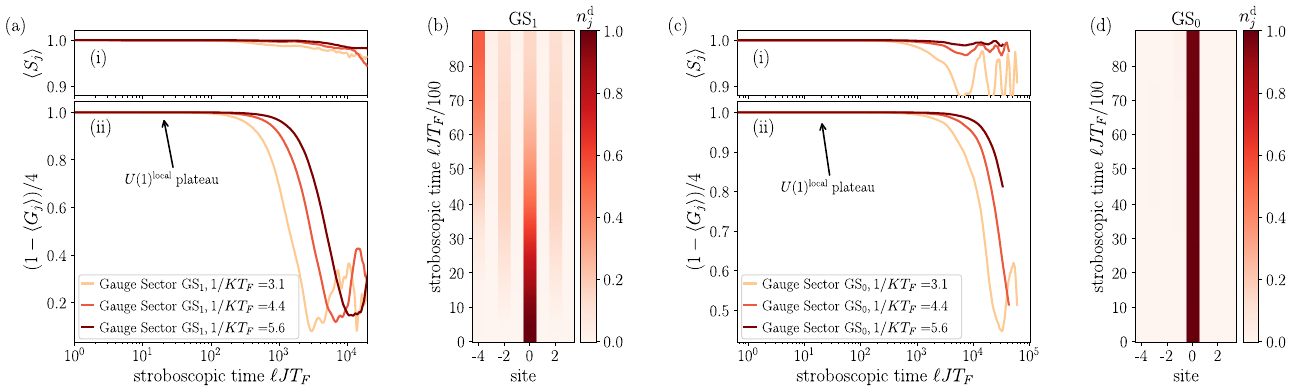}
    \caption{Dynamical protection of matter and gauge field dynamics with a spin-1 gauge field. (a) Dynamics of the local charge at site $j=0$ for the initial sector with one defect, $\text{GS}_1:\{g_j\} = \{-3, 1, 1, \cdots, 1\}$. The inserted figure: dynamics of (i) $\mathbb{Z}_2^\text{local}$ symmetry and (ii) the $U(1)^\text{local}$ symmetry violation. The prethermal lifetime of (i) is much longer than the lifetime of (ii). The defect spreads slowly, and the lifetime of the prethermal plateau can be controlled by the driving frequency.
    (b) Spatio-temporal distribution of the defect density $n^{\text{def}}_j = (1-\langle G_j\rangle)/4$ in the sector $\{g_j\} = \{-3, 1, 1, \cdots, 1\}$ with the driving frequency $1/KT_F = 5.6$. The defect remains largely localized at its initial position for long times. Moreover, due to the structure of the effective Hamiltonian, the motion of defects develops a density-wave pattern.   
    (c) Dynamics of the charge at site $j=0$ for the initial sector  $\text{GS}_0:\{g_j\} = \{1, 1, 1, \cdots, 1\}$. Unlike the case of spin-1/2, where the dynamics is totally frozen, the defect now can propagate, leading to the $U(1)^\text{local}$ symmetry violation. However, it can be seen that the dynamics remain sufficiently suppressed and the lifetime of the prethermal plateau is significantly longer than that in (a).
    (d) shows the spatio-temporal distribution of defect density $n^{\text{def}}_j = (1-\langle G_j\rangle)/4$ for the sector $\{g_j\} = \{1, 1, 1, \cdots, 1\}$ with the driving frequency $1/KT_F = 5.6$. The Defects are essentially frozen in their initial positions.
    The parameters in the simulation are $J = 1.0$, $K = 4.0,  h = 0.5, \epsilon_1 = \epsilon_2 = 1.0$. System size is $L = 10$.
    }
    \label{fig:spin1_SM}
\end{figure*}

\section{Generalization to arbitrary spin-S gauge fields}
\label{sec: spin-S}

Our results can be generalized to the case of a general spin-$S$ gauge field. In the general case, the local charge can take  eigenvalues of $g_j = -4S-1, -4S+1, -4S+3, \cdots, 4S+1$, while the defect-free sector is still chosen as $g_j=1, \; \forall j$. Here, the multiple of 4 is chosen so that when $S=1/2$, the eigenvalues of charge reduce back to $g_j = -3, -1, 1, 3$, which is consistent with our discussion of the spin-1/2 case.
Although general spin $S$ has a large local Hilbert-space dimension, the key idea that the subsymmetry $\mathbb{Z}_2^{\text{local}}$ induces selection rules and constrained dynamics that protect the target LGT, still works. 

For simplicity, we focus on the dynamics of the spin-$1$ gauge field case and initialize the system in the gauge-invariant  sector.
For one fixed site $j$, notice that $\mathbb{Z}_2^{\text{local}}$ preserving operations can induce transitions only within the following set of sectors
\begin{equation}
    g_j = (-3, 1, 5) \quad \text{and}\quad  g_j = (-5, -1, 3),
\end{equation}
because the spin configuration in each set possesses the same eigenvalues of the $S_j$ operator from Eq.~\eqref{eq:Sj}. Since here we discuss the system initialized to a state within the defect-free sector ($g_j=1, \forall j$), we need to only consider the first set $g_j = (-3, 1, 5)$.

The $U(1)^{\text{global}}$ symmetry ensures the conservation of total charge $\sum_jg_j$ so that the charge dynamics between two neighboring sites are
\begin{equation}
    \{g_j,g_{j+1}\} = \{1,1\}\leftrightarrow\{-3, 5\} \quad \text{or}\quad \{1,1\}\leftrightarrow\{5, -3\} \quad \text{or}\quad \{-3,1\}\leftrightarrow\{1, -3\} \quad \text{or}\quad \{1,5\}\leftrightarrow\{5, 1\}.
\end{equation}
Notice that, unlike the spin-$\frac{1}{2}$ case, cf.~case 1a in Sec.~\ref{sec.selectiontrule}, in the defect-free sector, different sites can now exchange $\Delta g=4$, since the upper limit for the charge eigenvalue now is 5. 

To enable an intuitive discussion of the possible dynamics, we present configurations in different $g_j$:
\begin{equation}
    g_j=1: \;^{0}\;_{\Uparrow}\;^{2},\;\;^{-2}\;_{\Uparrow}\;^{0},\;\;^{-2}\;_{\Downarrow}\;^{-2},\;\;^{0}\;_{\Downarrow}\;^{0}, \;\;^{2}\;_{\Downarrow}\;^{2}\;;
    \qquad 
    g_j=-3: \;^{0}\;_{\Uparrow}\;^{-2},\;\;^{2}\;_{\Uparrow}\;^{0},\; \;^{2}\;_{\Downarrow}\;^{-2}\;;
    \qquad 
    g_j=5:\;^{-2}\;_{\Downarrow}\;^{2}.
\end{equation}
Configurations with $g_j = -3,5$ are now defects, while configurations with $g_j = 1$ and spin-up matter fields ($\;^{0}\;_{\Uparrow}\;^{2},\;\;^{-2}\;_{\Uparrow}\;^{0}$) are termed kinks.

\begin{figure}[t]
    \centering
    \includegraphics[width=0.4\linewidth]{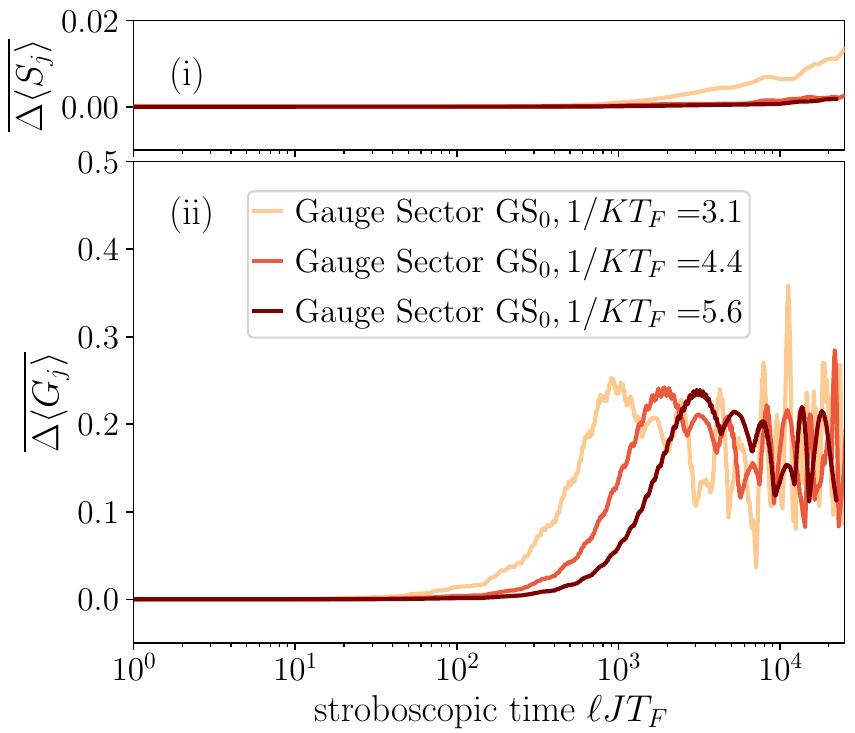}
    \caption{Dynamics of gauge violation with spin-1 gauge field for the physical sector $\text{GS}_p:\{g_j\}=\{1,-1,1,-1,1,-1,\cdots\}$ with open boundary condition. The inserted figure: dynamics of (i) $\mathbb{Z}_2^\text{local}$ symmetry and (ii) the $U(1)^\text{local}$ symmetry violation for the  physical sector.
    The violation is defined as $\overline{\Delta\langle S_j\rangle}=\frac{1}{L}\sum_{j=0}^{L-1}|\langle S_j\rangle-(-1)^{j+1}|$ and $\overline{\Delta\langle G_j\rangle}=\frac{1}{L}\sum_{j=0}^{L-1}|\langle G_j\rangle-(-1)^{j}|$.
    Compared to the spin-1/2 case, the degree of gauge violation within the system is greater at the same time scale, but it's still largely suppressed by the emergent local $\mathbb{Z}_2$ symmetry and the lifetime of the prethermal plateau can be controlled by the driving frequency.
    After undergoing a certain degree of gauge violation, the system enter another plateau, suggesting that on longer timescales, we can still obtain dynamics that are well approximated by exact LGT dynamics.
    The parameters in the simulation are $J = 1.0$, $K = 4.0,  h = 0.5, \epsilon_1 = \epsilon_2 = 1.0$. System size is $L = 8$.}
    \label{fig:sm_spin_1_phy}
\end{figure}

To compare with the dynamics of the system with spin-$\frac{1}{2}$ gauge field, we will discuss the dynamics of the effective Hamiltonian with the same driving protocol Eq.~\eqref{eq:dirve_simple} in the main text.
\begin{eqnarray}
Q_F^{(0)}{=}JH_{\text{LGT}}, \quad
Q_F^{(1)}{=}i\lambda_0[H_{\text{LGT}}, H_1],\quad \lambda_0 {=} \frac{KJ(J{+}K)}{2(2K{+}J)}T_F.
\end{eqnarray}
Here $H_{\text{LGT}} = \sum_{j=1}^L \sigma_j^+\tau_{j,j+1}^+\sigma_{j+1}^-+h.c.$, $H_1 = \sum_j \sigma_j^+\tau_{j,j+1}^x\sigma_{j+1}^-+h.c.$, where the operator $\tau_{j,j+1}^{+(-)}$ now becomes the raising (lowering) operator of spin-$S$. The dynamics of defects now allow for two types of evolution:
\begin{itemize}
    \item[(1)] The $g_j=5$ defects are totally frozen if there are no $g_j=-3$ defects nearby, while the  $g_j=-3$ evolves in the same way as in Fig.~\ref{fig:graprep}(c) with the effective QMM description we discussed above -- namely, that the defects can only hop when they collide with kinks. Here, we show some typical dynamics induced by the first-order effective Hamiltonian:
    \begin{eqnarray}
    &(i)\qquad &\;^{-2}\;_{\Downarrow}\;\underbrace{^{-2}\;_{\Downarrow}\;^{2}}_{g_j=5}\;_{\Downarrow}\;^{2} \; \text{    -- frozen dynamics of defect }g_j=5;\nonumber\\
    &(ii)\qquad &\;\underbrace{^{0}\;_{\Uparrow}\;^{-2}}_{g_j=-3}\;_{\Uparrow}\;^0\;_{\Downarrow}\;^0\;\rightleftharpoons\; \;^{0}\;_{\Downarrow}\;\underbrace{^{0}\;_{\Uparrow}\;^{-2}}_{g_j=-3}\;_{\Uparrow}\;^0
    \;\text{  -- typical QMM dynamics of defect }g_j=-3;
    \end{eqnarray} 
    \item[(2)] One can annihilate the kink configuration and create a pair of defects $g_j=-3$ and $g_j=5$. The reverse process can also occur. The typical dynamic is
    \begin{eqnarray}
        (iii)\qquad  \;\underbrace{^{-2}\;_{\Downarrow}\;}_{g_j=5}\underbrace{^{2}\;_{\Uparrow}\;^{0}}_{g_j=-3} \;_{\Uparrow}\;^{2}
    \;\rightleftharpoons \;\;^{-2}\;_{\Uparrow}\;^0\;_{\Uparrow}\;^{2}\;_{\Downarrow}\;^{2}\;\text{  -- creation and annihilation of the }g_j=-3,5\text{ pair}.
    \end{eqnarray}
\end{itemize}
However, both types of defect dynamics depend on the motion of kink configurations $g_j=1: \;^{0}\;_{\Uparrow}\;^{2},\;\;^{-2}\;_{\Uparrow}\;^{0}$; therefore, our previous discussion on the suppression of defect dynamics and its dependence on the energy degeneracy structures still holds.

To validate our above discussion, we performed numerical simulations on a system with a spin-1 gauge field. The results for the sectors $\text{GS}_1:\{g_j\} = \{-3, 1, 1, \cdots, 1\}$ and $\text{GS}_0:\{g_j\} = \{1, 1, 1, \cdots, 1\}$ are shown in Fig.~\ref{fig:spin1_SM} (a), (b) and Fig.~\ref{fig:spin1_SM} (c), (d) respectively. From the results we can observe the emergent hierarchical symmetry breaking $U(1)^\text{local} \to \mathbb{Z}^\text{local}_2 \times U(1)^\text{global}$, manifested in the enhanced increase in lifetime of the $\mathbb{Z}^\text{local}_2$-prethermal plateau as compared to the $U(1)^\text{local}$-plateau.
In Fig.~\ref{fig:spin1_SM} (c), we show the dynamics of a system initialized in the sector $\text{GS}_0$. In the main text we show the dynamics with the initial state in the sector $\text{GS}_0$ in Fig.~\ref{fig:floq}, where the defect can not be created under $\mathbb{Z}_2^{\text{local}}$-preserving perturbation due to the selection rule in Eq.~\eqref{eq.twositesrule}. In the spin-1 case, even though the selection rules we discussed above allow defects to be created and propagate, leading to the $U(1)^\text{local}$-symmetry violation. However, the numerical results in Fig.~\ref{fig:spin1_SM} (c) show the $U(1)^{\text{local}}$ plateau has a lifetime over $10^4$ even for a relatively small driving frequency, indicating that the dynamics of defects remain largely suppressed. 
Moreover, to demonstrate that the mechanism of gauge violation suppression from emergent symmetry also applies to physical sector, we present simulation results in Fig.~\ref{fig:sm_spin_1_phy}. Similar to the spin-1/2 case, we observe a relatively long preheating plateau whose lifetime can be tuned by adjusting the driving frequency. Furthermore, the system also enters a new plateau after a small gauge violation, suggesting that we can achieve dynamics well described by exact LGT over longer timescales through the driving protocol.

\end{document}